\documentclass[preprint,prb,showpacs,preprintnumbers,amsmath,amssymb]{revtex4}
\usepackage{dcolumn}
\usepackage{bm}
\usepackage{graphicx,epsfig}
\usepackage{bbm}
\usepackage[usenames]{color}
\usepackage{ulem}
\newcommand\beq{\begin{equation}}
\newcommand\eeq{\end{equation}}
\newcommand\bea{\begin{eqnarray}}
\newcommand\eea{\end{eqnarray}}

\newcommand\bra{\langle}
\newcommand\ket{\rangle}

\begin{document}

\title{{\bf Double Time Window Targeting Technique: Real-time DMRG dynamics in the PPP model}}

\author{Tirthankar Dutta and S.Ramasesha}

\email{tirthankar@sscu.iisc.ernet.in; ramasesh@sscu.iisc.ernet.in}

\affiliation{Solid State and Structural Chemistry Unit, Indian Institute of
Science, Bangalore 560012, India}

\begin{abstract}
We present a generalized adaptive time-dependent density matrix renormalization group (DMRG) scheme, called 
the {\it double time window targeting} (DTWT) technique, which gives accurate results with nominal 
computational resources, within reasonable computational time. This procedure originates from the 
amalgamation of the features of pace keeping DMRG algorithm, first proposed by Luo {\it et. al}, 
[Phys.Rev. Lett. {\bf 91}, 049701 (2003)], and the time-step targeting (TST) algorithm by Feiguin and
White [Phys. Rev. B {\bf 72}, 020404 (2005)]. Using the DTWT technique, we study the 
phenomena of spin-charge separation in conjugated polymers (materials for molecular electronics and 
spintronics), which have long-range electron-electron interactions and belong to the class of strongly 
correlated low-dimensional many-body systems. The issue of real time dynamics within  the Pariser-Parr-Pople 
(PPP) model which includes long-range electron correlations has not been addressed in the literature so far. 
The present study on PPP chains has revealed that, (i) long-range electron correlations enable both the 
charge and spin degree of freedom of the electron, to propagate faster in the PPP model compared to Hubbard 
model, (ii) for standard parameters of the PPP model as applied to conjugated polymers, the charge velocity 
is almost twice that of the spin velocity and, (iii) the simplistic interpretation of long-range correlations 
by merely renormalizing the {\it U} value of the Hubbard model fails to explain the dynamics of doped 
holes/electrons in the PPP model.
\end{abstract}

\pacs{73. 63 Nm}

\maketitle

\section{INTRODUCTION}
Low-dimensional strongly correlated many-body systems have been in the fore-front of both theoretical
and experimental researches for many decades, owing to their physics being very different from that of 
the three-dimensional systems. For example, these materials exhibit the phenomena of spin-charge separation, 
wherein the spin and charge degrees of freedom of the electron get decoupled from each other and propagate 
independently with different velocities. From an application point of view also, these materials are in
enormous demand. Amongs the genre of low-dimensional strongly correlated materials, the $\pi$-conjugated
polymers have attracted huge interest, being potential candidates for various molecular electronic and 
spintronic applications; examples include the organic light emitting diodes, organic semiconductors and 
organic thin-film transistors \cite{torsi,katz,marks,nitzan}. However, spin and charge transport in these
materials is still far from well understood because of the strong long-range electron-electron correlations
that exist in these systems. Transport is strictly a non-equilibrium phenomena, to understand which, one 
needs to investigate time evolution of appropriate wave packets in these strongly correlated low-dimensional 
systems. The density matrix renormalization group (DMRG) technique advanced by Steve White 
\cite{white1,peschel}, has proved to be a very powerful numerical method for studying large interacting 
low-dimensional quantum lattice systems. Originally formulated as a ground state method, this technique has 
been mostly used for studying static (equilibrium) quantum many-particle phenomena. Later, it was extended 
to calculate frequency-dependent spectral functions by the correction vector or the Lanczos techniques 
\cite{rama1,rama2,rama3,rama4,karen,white2,ul1},\cite{peter1}, thereby adapting it to deal with dynamical 
(equilibrium) many-body phenomena. In this regard, the dynamical DMRG (DDMRG) method \cite{jekel} turned out 
to be the spectral method, best suited for obtaining extremely accurate spectra. However, the DDMRG 
technique is limited to only one momentum and a narrow frequency range at a time. Constructing an entire 
spectrum using this method in order to obtain a reasonable grid in frequency and momentum space involves
independent runs for each frequency and momentum and is therefore, computationally highly intensive. 
An alternative route exists for obtaining an entire spectrum in a single run which involves time
evolving an appropriate wave packet in real space and time, followed by converting the information from 
($\vec{r}$,$t$) space to ($\vec{k}$,$\omega$) space using double Fourier transform. Quantum dynamics of a 
wave packet in real space time is needed to obtain the ($\vec{r}$,$t$) data, which for large systems was not 
feasible, until recently. Three of the recent time-dependent DMRG (t-DMRG) techniques are the pace keeping 
DMRG scheme due to Luo, Xiang and Wang (LXW) \cite{xiang}, the adaptive DMRG (t-DMRG) method 
\cite{vidal,daley,white3} and, the time-step targeting (TST) technique due to White \cite{white4}. 
In this paper we present a time-dependent DMRG scheme which is a hybrid of the LXW and TST algorithms. The 
organization of the paper is as follows: In Sec. I we give an overview of the existing 
time-dependent DMRG techniques, followed by a detailed discussion of the LXW and TST algorithms in order to 
compare their strengths and weaknesses. Section II is devoted to a detailed presentation of our algorithm. 
Section III  discusses some numerical issues associated with the DTWT scheme. In Sec. IV we compare the 
technique with the LXW and TST schemes. In Sec. V we present real time dynamics of spin-charge 
separation in regular Pariser-Parr-Pople (PPP) chains using our DTWT algorithm. Section VI provides the 
summary and conclusions of this work.

\subsection{Overview of existing time-dependent DMRG methods}
The study of out-of-equilibrium phenomena in strongly correlated low-dimensional systems has attained a 
forefront; DMRG has played a significant role in this context. Real space-time quantum dynamics using DMRG 
was introduced by Cazalilla and Marston \cite{cazalilla}. They calculated time evolution of one-dimensional 
systems under the influence of an applied bias. Their approach involved the use of DMRG for constructing the 
Hilbert space of the Hamiltonian using only the ground state, and subsequently solving the time-dependent 
Schr\"odinger equation numerically using the Hamiltonian matrix obtained in a {\it fixed basis}. Hence, the 
approach of Cazalilla and Marston is essentially static with respect to the Hilbert space in which time 
evolution is carried out. It is expected that when the evolving wave function becomes significantly different 
from the ground state, i.e., ''it moves out of the Hilbert space used for time evolution,'' it will loose 
accuracy. However, in the systems studied by them, this (expected) loss in accuracy with time did not occur 
within the time period for which they carried out the time evolution. The drawback of performing time 
evolution using a fixed DMRG basis is, one needs to keep a substantially large number of the density 
matrix eigenvectors (DMEVs), {\it m}, so that the evolved state is well described by the DMRG basis at large 
times. Luo, Xiang and Wang \cite{xiang} showed an effective way, also known as pace keeping DMRG, to 
construct DMEV basis for time evolving a wave function over a longer time interval. This is done by 
constructing a weighted average density matrix from the time evolved wave functions at discrete time steps 
$\Delta \tau$ in the time interval {\it 0-T}. Thus,
\beq
\rho_{L/R} ~=~ Tr_{R/L} \sum_{i=0}^{N_{t}} \alpha_{i} \mid \psi(t_{i}) \ket \bra \psi(t_{i}) \mid; ~~\sum_{i} \alpha_{i} = 1, 
\eeq 
where $N_{t}$ = $\frac{T}{\Delta \tau}$. The weights $\alpha_{i}$ are taken to be 1/2 for i = 0 and  
$\frac{1}{2N_{t}}$ otherwise. However, their approach suffers from two significant drawbacks. The LXW scheme 
performs {\it full} time evolution of the wave function at each infinite DMRG step thereby making it 
extremely time consuming. In DMRG calculations with multiple target states, truncation error is reduced by 
keeping the number of DMEVs ({\it m}), greater than the number of target states. But in the LXW algorithm, 
the number of target states $N_{t}$ [each target state corresponding to $\psi(t_{i})$  at time $t_{i}$] is 
usually $>>$ {\it m}, thereby decreasing accuracy. Modification of the original LXW scheme
in which $\rho_{L/R}$ is obtained as an average in each sub-interval $\Delta t$, where $\Delta t$ = 
$p\Delta \tau$, $p << ~N_{t}$ leads to a decrease in number of target states required, and consequently 
increases accuracy of the results \cite{dutta}. In an earlier study, we demonstrated this for {\it T} = 33 
fs, $N_{t}$ = 50,000, $\Delta \tau$ = 0.00066 fs and {\it p} = 200-500. In fact, the accuracy is not 
degraded even for {\it p} = 100. However, our earlier work has shown that this modification works well for 
nearest-neighbor Hamiltonians like the tight-binding (H\"uckel) and Hubbard models only. But, in
the case of models with long-range interactions, like the PPP Hamiltonian, the above modification fails and 
we need to average the density matrix over all time steps. 

In the context of simulating time evolution of matrix product states (MPS), Vidal developed a novel 
numerical scheme called the time-evolving block decimation technique \cite{vidal}. As DMRG is closely related
to MPS, this method was immediately utilized by the DMRG community to develop a very powerful numerical 
technique called adaptive t-DMRG \cite{daley,white3}, for studying real-time quantum dynamics in strongly 
correlated many-particle systems. The key idea of t-DMRG technique is to incorporate the Suzuki-Trotter (ST) 
decomposition of the time evolution operator exp($-i \Delta \tau H$) into the finite DMRG algorithm. Usually,
second-order ST decomposition is used but higher order ST decompositions of the matrix exponential
can be employed as well. The second-order Suzuki-Trotter breakup is
\beq
e^{-i \Delta \tau H} ~\approx~ e^{-i \Delta \tau H_{A}/2} e^{-i \Delta \tau H_{B}} e^{-i \Delta \tau H_{A}/2} ~+~ O(\Delta \tau^{3}),
\eeq
where $H_{A}$ is the Hamiltonian for the bond connecting sites {\it (2j-1, 2j)} and $H_{B}$ connects sites
{\it (2j, 2j+1)}; {\it j} = 1,2 $\cdots$ $\frac{N}{2}$, for $H_{A}$ and {\it j} =1,2, $\cdots$ 
$\frac{N}{2}$-1, for $H_{B}$. This breakup leads to grouping of terms such that all terms within $H_{A}$ or 
within $H_{B}$ which commute with each other, leading to high accuracy in the Trotter breakup. A DMRG 
super-block state, at a particular step {\it n} of the finite-system sweep, can be represented as
\beq
\mid \psi \ket ~=~ \sum_{L} \sum_{\alpha_{n},\alpha_{n+1}} \sum_{R} \psi_{L \alpha_{n} \alpha_{n+1} R} \mid L \ket \mid \alpha_{n} \alpha_{n+1} \ket \mid R \ket,
\eeq
where $\mid L \ket$ and $\mid R \ket$ represent the truncated DMEV basis states of the left and right 
blocks, and $\mid \alpha_{n} \alpha_{n+1} \ket$ represent the Fock-space states for the two central sites. 
Thus, an operator ($\hat{A}$) acting on the two central sites can be exactly expressed in  the same optimal 
basis as
\beq
[\hat{A} \psi]_{L \alpha_{n} \alpha_{n+1} R} ~=~ \sum_{\alpha_{n}^{'} \alpha_{n+1}^{'}} \hat{A}_{\alpha_{n} \alpha_{n+1} ; \alpha_{n}^{'} \alpha_{n+1}^{'}} \psi_{L \alpha_{n}^{'} \alpha_{n+1}^{'} R}.
\eeq
The ST decomposed time evolution operator for the bond connecting sites {\it n} and {\it n+1} can be applied 
exactly on the DMRG super-block state at step {\it n} of the finite DMRG algorithm. A full finite DMRG sweep 
corresponds to a time evolution by 2$\Delta \tau$ of the full system. In the t-DMRG scheme, two types of 
errors are involved: (1) the DMRG truncation error and (2) the ST decomposition error, which for second-order
decomposition is ($O\Delta \tau^{3})$, in each time step. Adaptive t-DMRG based on the Suzuki-Trotter 
decompositions of the time evolution operator is restricted to nearest-neighbor interactions only since the 
break-up into $H_{A}$ and $H_{B}$ is valid only in this case. Schmitteckert \cite{peter2} proposed a 
Krylov space approach for obtaining the exponential time evolution operator. In this method a small matrix 
representation of the original Hamiltonian, {\it H}, is obtained in the basis vectors of the Krylov space, 
namely $\{\overrightarrow{x}, H\overrightarrow{x}, H^{2}\overrightarrow{x}, \cdots, H^{l-1}\overrightarrow{x}
; ~\overrightarrow{x}, \text{an arbitrary vector}\}$ for an {\it lth} dimensional Krylov space. This permits 
explicit form of the time evolution operator in a finite optimized basis. He, however, employed the LXW 
scheme for obtaining the DMRG space spanned by the Hamiltonian.

White proposed a second adaptive approach called the time-step targeting (TST) technique to circumvent 
the nearest-neighbor limitation associated with ST break-up \cite{white4}. Two key ideas are introduced in 
this scheme. The first being, there is no conceptual need to have the {\it same} time step for time evolution
and for building the Hilbert space of the evolving wave function. This simply implies that the time-scales 
$\Delta \tau$ on which time evolution is discretized and $\Delta t$ on which the DMEV basis is adapted, need 
not be identical (this is the crucial difference between the LXW and TST schemes). The observation from the 
work of Cazalilla and Marston that even a static state space remains a good choice for some finite time, 
implies the choice of $\Delta t ~>~ \Delta \tau$. Thus, one can evolve the wave function using any of the 
non-Trotter methods, the one used by White and Feiguin being fourth-order Runge-Kutta (R-K) technique. Other 
available methods are the Crank-Nicholson scheme or Krylov-space-based decomposition of the matrix 
exponential \cite{noack}. The second key idea of the TST technique is to adaptively build the Hilbert space 
representing the time evolving wave function, at intervals $\Delta t$, from the states which are expected to 
appear in future through time evolution. For this purpose, several DMRG sweeps are carried out at a fixed 
time {\it t}. For each $(L-\alpha_{n}-\alpha_{n+1}-R)$ configuration during these finite sweeps, a 
fourth-order Runge-Kutta integration is carried out, and the time-evolved states 
$\mid \psi(t + j \Delta t) \ket$, $~j ~=~ 0, \frac{1}{3}, \frac{2}{3}, 1$ are used to build the reduced 
density matrices of the blocks using Eq. (1). After sufficient number of sweeps, when the Hilbert space 
optimally represents the wave function $\mid \psi(t) \ket$, final time evolution is performed using a time 
step $\Delta \tau ~\approx~ \frac{\Delta t}{10}$. The new wave function is then used to build the DMEV basis 
for the next time propagation. Both LXW and the TST techniques are {\it generalized} adaptive time-dependent 
DMRG schemes since they can be applied to any Hamiltonian, with arbitrary range (beyond nearest neighbor) of 
interactions. This flexibility stems from the fact that the time evolution operator is not decomposed
using Suzuki-Trotter-type decomposition schemes. In the LXW technique, adaptive construction of the Hilbert
space of the time-evolving initial state as well as temporal propagation are performed within the context 
of infinite DMRG algorithm. On the other hand, TST technique uses the finite DMRG scheme to adaptively 
update the Hilbert space of the time evolving wave function as well as, propagate it in time. The time steps 
used for state space updating and time evolution of the wave packet are the same in LXW scheme while they are 
chosen to be independent in the TST scheme. As stated by White, the time step for basis adaptation 
($\Delta t$) is larger than that for evolution ($\Delta \tau$). 

We have compared the two techniques and have found that, for a given value of {\it m} and system size 
{\it N}, LXW scheme besides being comparable in speed with the TST algorithm, is also more accurate. 
In the LXW scheme, at every system size we need to evolve the wave packet over the entire time period,
before getting the DMEV basis for moving to the next system size. In the TST scheme, although we do the
time evolution only for the desired finite-system size, we employ a finite DMRG scheme at every time
evolution step $\Delta \tau$, which entails large CPU times. The higher accuracy of the LXW scheme arises 
from the fact that DMEV basis is obtained from a weighted average density matrix, constructed from 
time-evolved wave packets at all time intervals, while in TST scheme the weighted average density matrix is 
constructed from the wave packets 0, $\frac{\Delta t}{3}$, $\frac{2 \Delta t}{3}$, and $\Delta t$ for each 
time step of evolution $\Delta \tau$.

If one uses the {\it window modification} of the LXW technique as proposed by us \cite{dutta}, thereby 
decreasing the number of target states, the accuracy as well as computational speed can be further improved. 
The above observations leads one to conclude that the LXW technique is superior compared to the TST scheme.
However, LXW scheme also suffers from two serious drawbacks: first, LXW method is not {\it quasiexact} 
unlike the TST scheme. An approximate DMRG algorithm is quasiexact when the error in the observables is 
strictly controlled by the truncation error, $\epsilon ~=~ 1.0 -\sum_{i} \rho_{i}$, $\rho_{i}$ being the 
dominant eigenvalues corresponding to the reduced density matrix eigenvectors which are retained. In case of 
a quasiexact scheme, the errors in expectation values of the system's properties are either proportional to 
$\epsilon$ or $\sqrt{\epsilon}$, that is, $(\bra \hat{O}(t) \ket_{DMRG} ~-~ \bra \hat{O}(t)\ket_{exact}) 
~\propto$ $\epsilon$ or $\sqrt{\epsilon}$. The infinite-system DMRG technique applied to a finite system is 
not a quasiexact approximation scheme even though $\epsilon$ goes to zero as {\it m} increases. However, in 
the absence of any metastable ground states and with ''sufficient'' number of sweeps, the finite system DMRG 
algorithm is quasiexact. LXW scheme being an infinite system DMRG algorithm applied to finite systems, is  
nonquasiexact \cite{white4}. Second, in case of long-range Coulomb interaction as in the PPP Hamiltonian, 
our {\it window modification} fails, thereby making it necessary to retain all the time-dependent target 
states. This makes the LXW scheme computationally inefficient. Finally, both the LXW and TST techniques are 
computationally time consuming. These observations motivated us to develop a real-time evolution method 
which will have the strengths of both these methods while overcoming their limitations of long computational 
times and poor accuracy, especially for long-range interacting models.

\section{DOUBLE TIME WINDOW TARGETING (DTWT) ALGORITHM}

Conceptually, the scheme which we have implemented is a hybrid of the LXW and TST algorithms 
described in the previous section. These schemes differ from each other mainly in the prescription for 
constructing the weighted average density matrix. The different schemes are compared schematically in Fig. 1.
In the LXW scheme, the weighted average density matrix is obtained from the density matrices of the time 
evolved states at all times ({\it 0}, {\it T}) in steps of $\Delta \tau$, the total time interval from 
{\it 0} $\rightarrow$ {\it T} being $\Delta t$. In the TST scheme, in each time interval $\Delta t$, a 
weighted average density matrix is constructed, typically with the states at the beginning and end
of the time interval and at two equidistant intermediate points. The time evolution is carried out with a
time step $\Delta \tau$ such that, $\Delta \tau$ is $\sim$ $\frac{\Delta t}{10}$ within the interval.
For time evolution over the next time step $\Delta \tau$ to 2$\Delta \tau$, the density matrix is
constructed using the time interval from $\Delta \tau$ to ($\Delta t$ + $\Delta \tau$). Besides, in the
TST algorithm, unlike in the LXW algorithm, finite DMRG procedure is carried out at each time at which the 
density matrix computation is carried out. In the present scheme which we call the DTWT technique, we 
construct the density matrix over a time interval $2p\Delta \tau$, as a weighted average of the density 
matrices built at time steps of length $\Delta \tau$, and use the DMEVs from the resulting density matrix to 
constructing the Hilbert space of the Hamiltonian, for time evolving the desired wave packet from 
$\Delta \tau$ to $p\Delta \tau$. In this technique, each time interval of length $\Delta t$ is broken into 
{\it p} sub-intervals of length $\Delta \tau$ such that, while the wave packet evolves by $\Delta t$, the 
basis is adapted over an interval 2$\Delta t$. After every time evolution by $\Delta t$, the interval is 
slided by 2$\Delta t$ for constructing the Hilbert space for the next time evolution. Thus, approximate
future states for a time period 2$\Delta t$ are used for evolving the wave packet over a time 
interval $\Delta t$, in steps of $\Delta \tau$. As in the TST scheme, we employ finite procedure to 
get accurate wave functions in each time interval.
\begin{figure}
\begin{center}
\epsfig{file=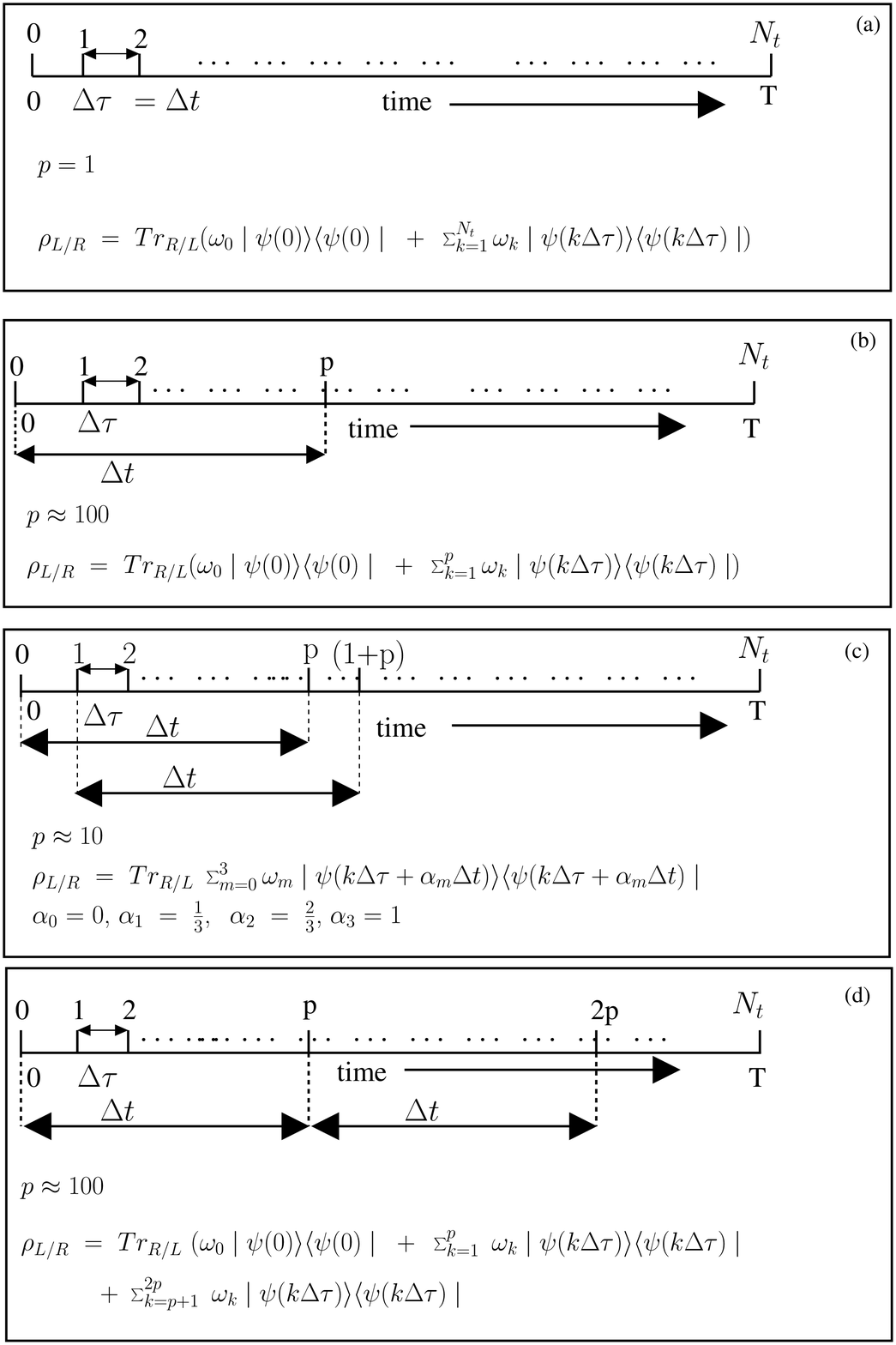,width=4.5in}
\caption{Pictorial representation of the construction of the time averaged reduced density matrices for 
left/right block ($\rho_{L/R}$) in the (a) original LXW, (b) modified LXW, (c) TST, and (d) DTWT schemes. 
In all the schemes, $\Delta \tau ~=~ \frac{T}{N_{t}}$ and $\Delta t ~=~ p\Delta \tau$. In the original LXW 
scheme (a), $\Delta t$ = $\Delta \tau$ ($p$ = 1) is used while in our modified LXW scheme (b), 
$\Delta t ~=~ p\Delta \tau$, with $p \approx 100$ is used. In the TST scheme (c), a sliding time window of 
length $\Delta t$ is used for updating time step $\Delta \tau$ ($\Delta t ~\approx 10 \Delta \tau$). In the 
DTWT method (d), a sliding time window of length 2$\Delta t$ = $2p\Delta \tau$, is used for updating time 
window of length $\Delta t$ = $p\Delta \tau$, $p ~\approx$ 100. In the last two schemes, finite DMRG 
procedure is carried out to obtain basis adaptation and time evolution.}
\end{center}
\end{figure}

\subsection{Initial wave-packet construction}

A conventional infinite DMRG algorithm is carried out to build the system of desired final size on which
time evolution is wished to be performed. We wish to study the dynamics of spin and charge transport in
a one-dimensional model with long-range interactions. Therefore, the initial wave packet at $t$ = 0 is 
formed by annihilating an upspin ($\uparrow$) electron from site $1$ of a neutral system in the ground 
state ($\mid \psi_{gs}^{0} 
\ket$),
\beq
\mid \psi(0) \ket ~=~ a_{1,\uparrow} \mid \psi_{gs}^{0} \ket.
\eeq
Since we are interested in the time evolution of this charged wave packet formed by annihilating an 
electron in the chain, the Hilbert space formed from the DMEVs of the neutral system alone, would be 
inappropriate. Hence we construct the half-block density matrices at each system size as a weighted average 
of the density matrices of the desired wave packet, the neutral ground state and other relevant states. These 
density matrices correspond to system of the same size but with different particle numbers. Thus, at each 
step of the infinite scheme, the half-block reduced density matrices are formed in the following way:
\begin{gather}
\begin{split}
\rho_{L/R} &=~ Tr_{R/L} \biggl(\omega_{0} \mid \psi(0) \ket \bra \psi(0) \mid ~+~ \sum_{j=1}^{r} \omega_{j} \mid \psi_{j} \ket \bra \psi_{j} \mid \biggr) \\
           &=~ Tr_{R/L} \biggl(\omega_{0} \mid \psi(0) \ket \bra \psi(0) \mid ~+~ \omega_{gs}^{0} \mid \psi_{gs}^{0} \ket \bra \psi_{gs}^{0} \mid ~+~ \sum_{j=2}^{r} \omega_{j} \mid \psi_{j} \ket \bra \psi_{j} \mid \biggr), 
\end{split}
\end{gather}
where, $\mid \psi(0) \ket$ is the initial state with weight $\omega_{0}$, $\mid \psi_{j} \ket$ is the 
{\it j}th relevant state having weight $\omega_{j}$, and 
$\biggl(\omega_{0} ~+~ \sum_{j=1}^{r} ~\omega_{j} ~=~ 1 \biggr)$ and the number of relevant states including 
the neutral ground state is $r$. Other states which can be considered as relevant target states are the 
ground state of the ion, especially in case of inhomogeneous systems. From our earlier studies on the LXW 
algorithm we have found that for systems such as (CN)$_{x}$ and (PN)$_{x}$ consisting of more than one type 
of atom, it is necessary to target the corresponding cationic ($\mid \psi_{gs}^{+} \ket$) or anionic 
($\mid \psi_{gs}^{-} \ket$) ground states also, in addition to the initial wave packet and neutral ground 
state. Thus, in this case the density matrix is given by
\beq
\rho_{L/R} ~=~ Tr_{R/L} \biggl(\omega_{0} \mid \psi(0) \ket \bra \psi(0) \mid ~+~ \omega_{gs}^{0} \mid \psi_{gs}^{0} \ket \bra \psi_{gs}^{0} \mid ~+~ \omega_{gs}^{+} \mid \psi_{gs}^{+} \ket \bra \psi_{gs}^{+} \ket \biggr).
\eeq
An exhaustive analysis of the dependence of the charge densities and spin densities of the initial 
wave packet on the weights of the target states showed that the nature of the initial state 
does is not very sensitive to the weights. Hence, as in the LXW scheme, we 
have chosen $\omega_{0}$ = 0.8 and $\omega_{gs}^{0} ~=~ \omega_{gs}^{+}$ = 0.1 for all the systems
we have studied. 

The initial state for the final lattice, which is obtained using the infinite DMRG scheme, is numerically 
evolved in time using the time dependent Schr\"odinger equation (TDSE). The purpose of this time evolution is
to obtain a set of time-dependent states, with which the initial Hilbert space of the time-evolving wave 
packet can be constructed for starting the finite DMRG scheme. The TDSE has been solved numerically using 
the second-order multistep differencing scheme (MSD2) \cite{cakmak}. We settled on this scheme after trying
more accurate schemes such as fourth- and sixth-order multistep differencing schemes (MSD4 and MSD6),
\cite{ittika}, and the fourth-order R-K method. The final evolution step is the one that needs accurate 
integration of the TDSE. The MSD2 scheme is given by the following equation:
\beq
\mid \psi(t+m\Delta \tau) \ket = -2i\tilde{H}\mid \psi(t+[m-1]\Delta \tau) \ket + \mid \psi(t+[m-2]\Delta \tau) \ket + O(\Delta \tau^{3}); m \in [2,2p]
\eeq
The Hamiltonians used in the present study are time-independent and $\tilde{H} ~=~ (H^{+}-E^{+}_{0})$, 
where $H^{+}$ is the Hamiltonian of the positively charged system (cation) and $E^{+}_{0}$ is the ground 
state eigenvalue of $H^{+}$. The flow-chart for the initial infinite DMRG procedure is given in Fig. 2
\begin{figure}
\begin{center}
\epsfig{file=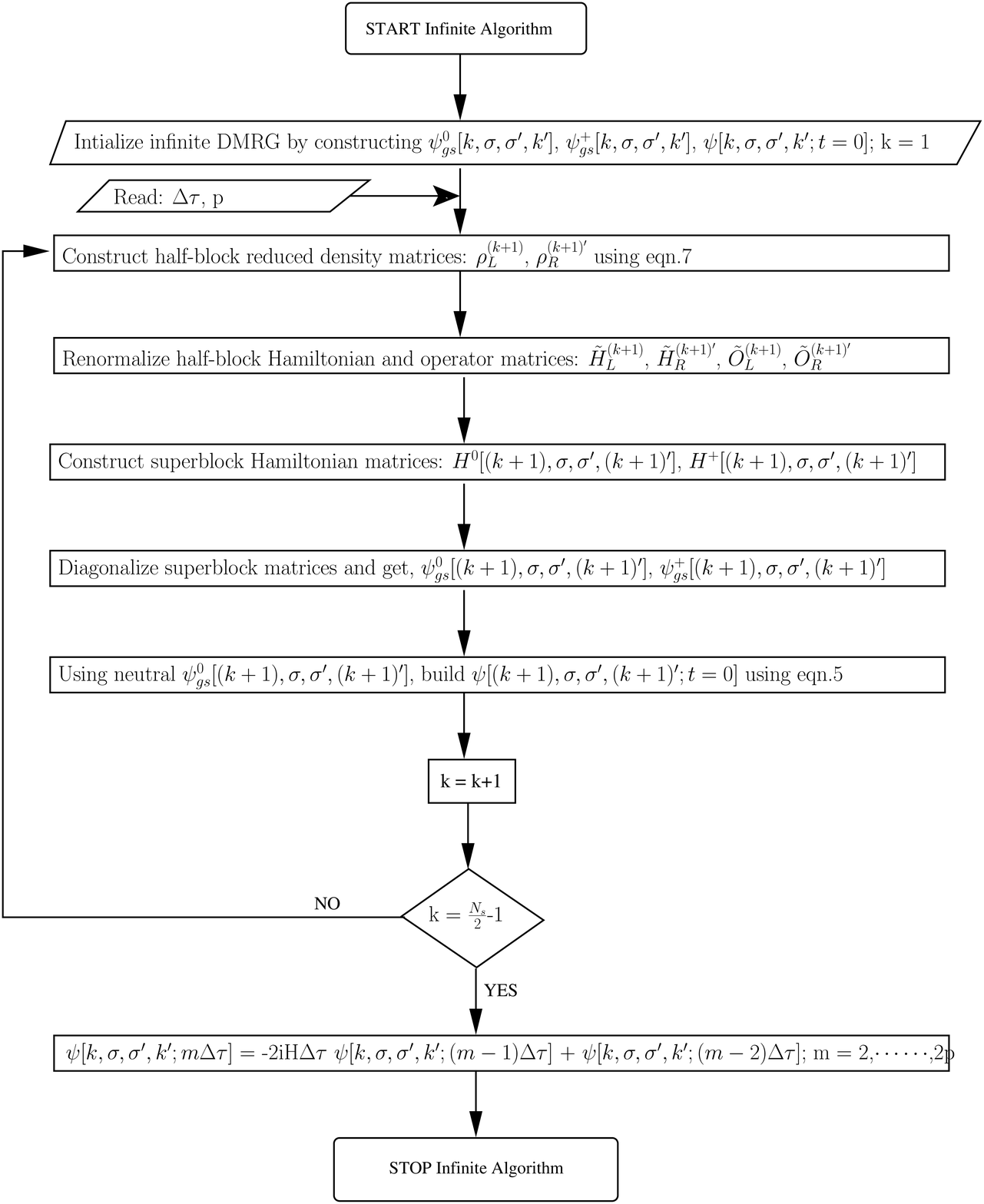,width=6.0 in}
\caption{Flow-chart for the infinite DMRG algorithm in the DTWT technique.  $k$ = 1,2,3,$\cdots ~\cdots$,
($\frac{N_{s}}{2}$-1), is the half-block length; $N_{s}$ is the total number of sites in the final lattice. 
$\sigma$ and $\sigma^{\prime}$ are the two new sites added during the infinite DMRG scheme; 
$\psi^{0}_{gs}[(k+1),\sigma,\sigma^{\prime},(k+1)^{\prime}]$ and $\psi^{+}_{gs}[(k+1),\sigma,\sigma^{\prime},
(k+1)^{\prime}]$ are the neutral and ionic ground states. $\psi[(k+1),\sigma,\sigma^{\prime},(k+1)^{\prime}; 
m\Delta \tau]$ is the time evolved wave packet at $t$ = $m\Delta \tau$, obtained using the MSD2 technique.}
\end{center}
\end{figure}

\subsection{Hilbert-space adaptation and time evolution using finite DMRG scheme}

After obtaining the initial wave packet, the neutral and the ionic ground states, and the time evolved wave 
packets $\mid \psi(k\Delta \tau) \ket$, $k \in [1, 2p]$, using infinite DMRG algorithm (see Fig. 2), for a 
system of desired size, finite DMRG procedure is carried out to adaptively construct the Hilbert space of the
Hamiltonian and propagate the wave function in time. Each finite DMRG step generates a system configuration 
having two blocks joined together by two single sites. Two full sweeps of the finite DMRG procedure 
involves 4$(\frac{N_{s}}{2}-2)$ steps, $N_{s}$ being the final system size; we call $(\frac{N_{s}}{2}-2)$ 
finite DMRG steps a half sweep. Hence, two full sweeps imply sweeping from 
($\frac{N_{s}}{2}-1,1,1,\frac{N_{s}}{2}-1$) $\rightarrow \cdots$ ($N_{s}-3,1,1,1$) $\rightarrow \cdots$ 
($\frac{N_{s}}{2}-1,1,1,\frac{N_{s}}{2}-1$) $\rightarrow \cdots$ (1,1,1,$N_{s}-3$) $\rightarrow \cdots$
($\frac{N_{s}}{2}-1,1,1,\frac{N_{s}}{2}-1$), where the first and last numbers in parenthesis give the
number of sites in the left and right blocks, respectively. At every step of the finite DMRG procedure 
starting from infinite DMRG solution corresponding to the system ($\frac{N_{s}}{2}-1,1,1,\frac{N_{s}}{2}-1$),
the following operations are performed:  

(1) At time $t$ = 0, the reduced density matrix  for the appropriate (left or right) block is computed using 
the neutral ground state $\mid \psi^{0}_{gs} \ket$ with an assigned weight $\omega_{gs}^{0}$, the initial 
wave packet $\mid \psi(0) \ket$ with weight $\omega_{0}$, the ionic ground state $\mid \psi^{+}_{gs} \ket$ 
with weight $\omega_{gs}^{+}$ and all other preliminary time evolved states $\mid \psi(k\Delta \tau) \ket$ 
({\it k} = 1,2,$\cdots$,2{\it p}) with weights $\omega_{k}$. The total weight is normalized to unity 
[Eq. (9)],
\beq
\begin{split}
\rho_{L/R} &=~ Tr_{R/L} \biggl(\omega_{gs}^{0} \mid \psi^{0}_{gs} \ket \bra \psi^{0}_{gs} \ket ~+~ \omega_{gs}^{+} \mid \psi_{gs}^{+} \ket \bra \psi_{gs}^{+} \mid  ~+~ \omega_{0} \mid \psi(0) \ket \bra \psi(0) \mid \\
           &+~ \sum_{k=1}^{2p} \omega_{k} \mid \psi(t+k\Delta \tau) \ket \bra \psi(t+k\Delta \tau) \mid \biggr).
\end{split}
\eeq
At other times ($t ~\ne$ 0), the reduced density matrix for the appropriate (left or right) block is 
constructed using the initial state, the ionic ground state, and 2$p\Delta \tau$ time-dependent states
(the neutral ground state is not considered), as given below:
\beq
\begin{split}
\rho_{L/R} &=~ Tr_{R/L} \biggl(\omega_{gs}^{+} \mid \psi_{gs}^{+} \ket \bra \psi_{gs}^{+} \mid  ~+~ \omega_{0} \mid \psi(0) \ket \bra \psi(0) \mid \\
           &+~ \sum_{k=1}^{2p} \omega_{k} \mid \psi(t+k\Delta \tau) \ket \bra \psi(t+k\Delta \tau) \mid \biggr).
\end{split}
\eeq
The weights $\omega_{gs}^{0}$, $\omega_{gs}^{+}$, and $\omega_{0}$ are adjusted (as in the infinite DMRG 
scheme used to generate $\mid \psi(t+k\Delta t) \ket$, {\it k} = 1,2,3,$\cdots \cdots$,2{\it p}, described 
in the previous section), after exhaustive analysis. In the case of TST algorithm, the weights of the target 
states can all be chosen to be equal or unequal. However, in our case we find that equal weightage of all the 
target states severely deteriorates the results. Our tests have shown that the optimal unequal weights for
Eq. (9) are $\omega_{0}$ = 0.7, and $\omega_{gs}^{0} ~=~ \omega_{gs}^{+}$ = 0.1 so that 
$\omega_{k} ~=~ \frac{(1.0 ~-~ \omega_{0} ~-~ \omega_{gs}^{0} ~-~ \omega_{gs}^{+})}{p}$ = $\frac{0.1}{p}$, 
$\forall ~k$. In case of Eq. (10), the optimal weights are $\omega_{0}$ = 0.8, $\omega_{gs}^{+}$ = 0.1, and 
$\omega_{k} ~=~ \frac{(1.0 ~-~ \omega_{0} ~-~ \omega_{gs}^{+})}{p}$ = $\frac{0.1}{p}$, $\forall ~k$. 

(2) The resulting left- or right-block density matrix is diagonalized and the DMEVs corresponding to the 
double time window, $t ~\rightarrow$ ($t+2p\Delta \tau$) are chosen. Using this DMEVs, the renormalized 
block Hamiltonians ($\tilde{H}_{L/R}$) and operators ($\tilde{O}_{L/R}$) are constructed. Using these 
renormalized Hamiltonians and operators, the super-block Hamiltonian for the current system configuration 
with required particle numbers is obtained.

(3) The super-block Hamiltonians for neutral and ionic systems are diagonalized using either the Lanczos 
\cite{lanczos}, Davidson's \cite{davidson}, or any other iterative sparse matrix diagonalization algorithm 
\cite{template} to obtain the ionic and neutral ground states. In case of the time evolution from $t$ = 0 to 
$t$ = $\Delta t$, both the neutral and the ionic ground states are needed while for the subsequent 
time evolutions ($t ~\ne$ 0), only the ionic ground state is required. This implies that two super-block 
diagonalizations are needed for the initial time evolution while only one super-block diagonalization for the
remaining evolutions, at every finite DMRG step. We have used the Davidson's algorithm for diagonalization of
the super-block Hamiltonian. 

(4) For time evolution from $t$ = 0 to $t$ = $\Delta t$ $\equiv$ $p \Delta \tau$, the initial wave packet is 
explicitly constructed from the neutral ground state according to Eq. (5). Hence, the neutral ground state 
is retained as a target state in Eq. (9). For the remaining time evolutions ($t ~\ne$ 0), the initial state 
is transformed from the old DMEV basis to the current one using White's wave function transformation 
\cite{wavetrans}; therefore the neutral ground state is no longer required.

(5) The initial wave packet and super-block Hamiltonian which are both expressed in the current DMEV 
basis, are used to solve the TDSE numerically using the MSD2 scheme from $t ~\rightarrow$ ($t+2\Delta t$) in 
time steps of $\Delta \tau$.
\vspace{0.5cm}

(6) After two full sweeps the final configuration ($\frac{N_{s}}{2}-1,1,1,\frac{N_{s}}{2}-1$) is evolved
from $t ~\rightarrow$ ($t+\Delta t$) $\equiv$ ($t+p\Delta \tau$). This is the final (single) time window 
evolution. This is carried out using the fourth-order R-K method. The fourth-order R-K method propagates 
$\mid \psi(t) \ket$ to $\mid \psi(t+\Delta \tau) \ket$ using four vectors $\mid k_{1} \ket$, 
$\mid k_{2} \ket$, $\mid k_{3} \ket$, and $\mid k_{4} \ket$ defined as
\begin{gather} 
\mid k_{1} \ket ~=~ -i\Delta \tau \tilde{H}(t) \mid \psi(t) \ket, \nonumber \\
\mid k_{2} \ket ~=~ -i\Delta \tau \tilde{H}(t + \frac{\Delta \tau}{2}) \biggl(\mid \psi(t) \ket ~+~ 1/2 \mid k_{1} \ket \biggr), \nonumber \\
\mid k_{3} \ket ~=~ -i\Delta \tau \tilde{H}(t + \frac{\Delta \tau}{2}) \biggl(\mid \psi(t) \ket ~+~ 1/2 \mid k_{2} \ket \biggr), \nonumber \\
\mid k_{4} \ket ~=~ -i\Delta \tau \tilde{H}(t + \Delta \tau) \biggl(\mid \psi(t) \ket ~+~ \mid k_{3} \ket \biggr). 
\end{gather}
{\noindent The state at time (t+$\Delta \tau$) is given by}
\beq
\mid \psi(t+\Delta \tau) \ket ~\approx ~\frac{1}{6} \biggl(\mid k_{1} \ket + 2\mid k_{2} \ket + 2\mid k_{3} \ket + \mid k_{4} \ket \biggr) + O(\Delta \tau^{5}).
\eeq 
The obtained time-dependent states can be either saved to disk for future use or they can be used on the fly 
for calculating dynamical observables. The final state $\mid \psi(t+p\Delta \tau) \ket$ is then used as the 
next initial state and the same procedure is repeated for the next single window time propagation (see 
Fig. 3). 

In the infinite DMRG part of the DTWT technique, the half-block density matrices and the corresponding
DMEVs are constructed from the initial wave packet, the neutral ground state and the ionic ground state, all 
of which are time-independent real wave functions. Hence in the infinite DMRG part of our algorithm, all 
operations are performed in real arithmetic and the quantities (scalars, vectors, and matrices) involved, are
all real. However, in the finite DMRG part we encounter time-dependent states which are complex quantities, 
resulting in complex reduced density matrices, DMEV basis, renormalized left- and right-block Hamiltonians, 
block operators, and super-block Hamiltonian matrices. Therefore, the DTWT method as implemented by us 
performs real arithmetic for the infinite DMRG part and complex arithmetic for the finite DMRG part.
\begin{figure}
\begin{center}
\epsfig{file=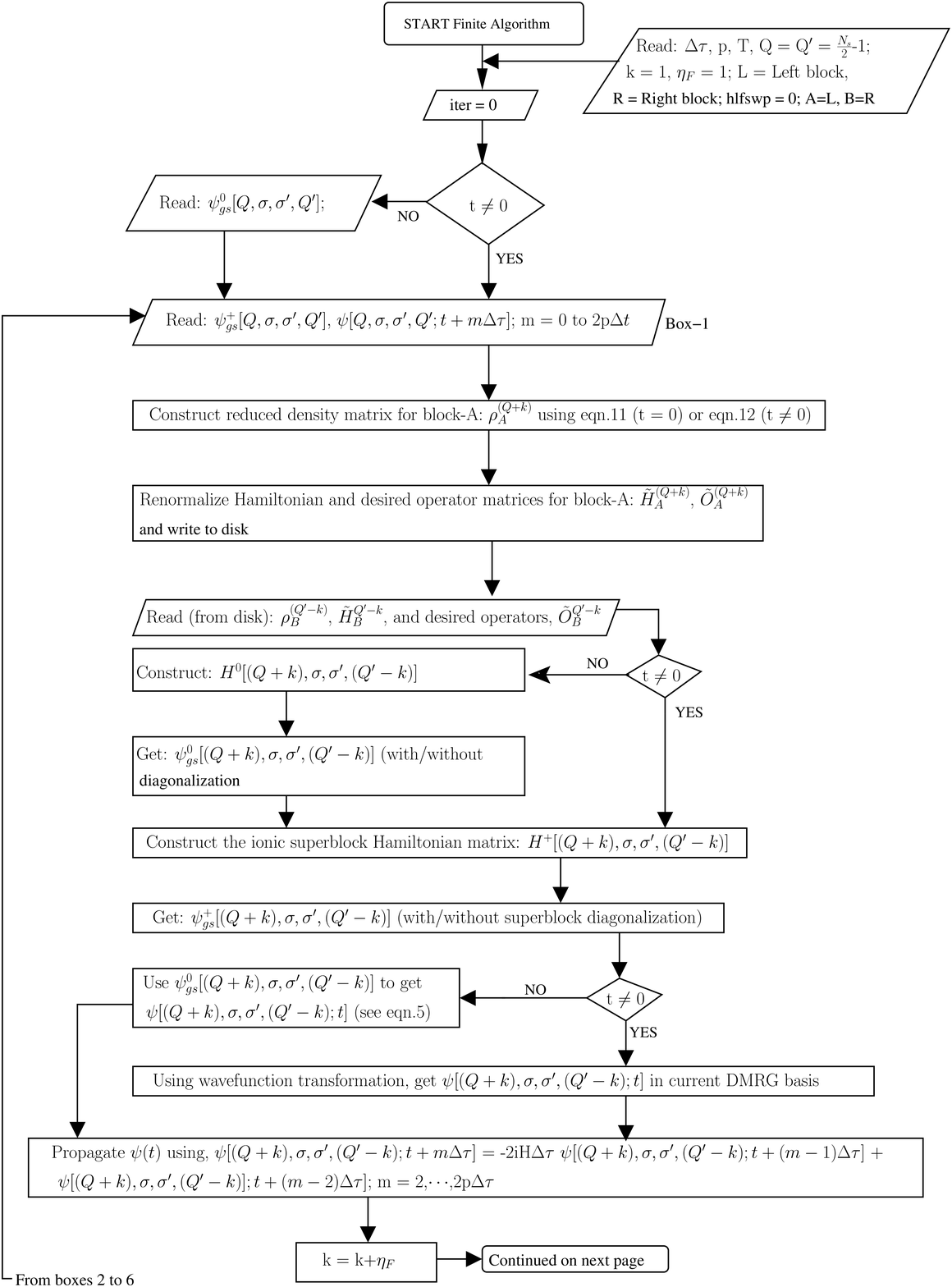,width=6.0 in} 
\end{center}
\end{figure}
\begin{figure}
\begin{center}
\epsfig{file=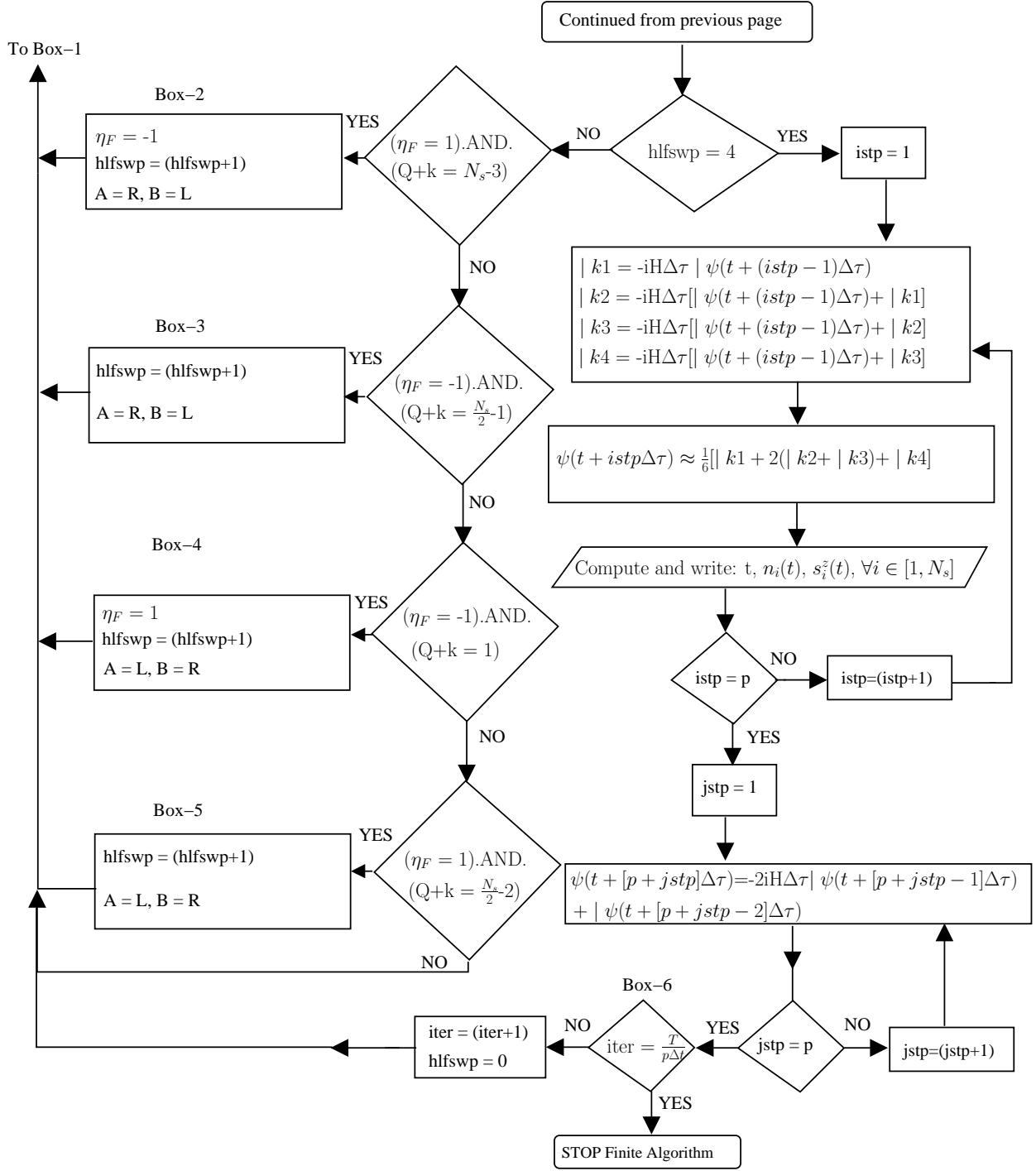,width=6.5in}
\caption{Basic scheme showing the use of finite DMRG algorithm in the DTWT technique. $Q+k$ ($Q'-k$) is the 
length of 
left (right) block. $\eta_{F}$ denotes the direction of sweep: $\eta_{F}$ = 1 (-1) implies left (right) 
$\rightarrow$ right (left) sweep. hlfswp = number of half sweeps. Time evolution is performed using 
fourth-order R-K technique while basis adaptation is done using the MSD2 technique.}
\end{center}
\end{figure}

\newpage
\section{SOME NUMERIAL ISSUES}

It is well known in the literature that the sparse super-block matrix diagonalization is the most 
time-consuming step in both the infinite and finite DMRG schemes. For the time evolution from $t$ = 0 to 
$t$ = $p\Delta \tau$, each step of the finite part of our DTWT algorithm involves two sparse matrix 
diagonalizations: one for the neutral system's Hamiltonian and the other for the ionic system's Hamiltonian.
For all subsequent time evolutions, $t ~>$ $p\Delta \tau$, diagonalization of only the super-block 
Hamiltonian for the ionic system is required at each step of the finite part of the DTWT procedure. If the 
sparse matrix diagonalizations involved at each step of the finite DMRG part of our algorithm can be replaced by the wavefunction transformation introduced by White \cite{wavetrans}, computational time is substantially 
reduced. However, our tests indicated that for the finite DMRG steps, ($N_{s}-4,1,1,2$) $\leftrightarrows$ 
($N_{s}-3,1,1,1$) and ($2,1,1,N_{s}-4$) $\leftrightarrows$ ($1,1,1,N_{s}-3$), the wave function 
transformation fails to be accurate. Thus, the matrix diagonalization step needs to be selectively replaced
by the wave function transformation. Taking recourse to White's wavefunction transformation, we reduce
the number of sparse matrix diagonalizations for the initial time evolution ({\it 0} $\rightarrow$ 
$p\Delta \tau$), from 8($\frac{N_{s}}{2}$-2) to 8 for two full sweeps; for all subsequent time evolutions 
($t ~>$ $p\Delta \tau$), the wave function transformation reduces the number of super-block Hamiltonian 
diagonalizations from 4($\frac{N_{s}}{2}$-2) to 4, for (every) two full sweeps, thereby reducing the 
computational time significantly. Obtaining the ground state wave function and energy through matrix 
diagonalization involves the solution of an eigenvalue equation, 
$\biggl(H \mid \psi_{gs} \ket ~=~ E_{gs} \mid \psi_{gs} \ket \biggr)$ while obtaining the same using the 
wave function transformation involves transformation of the ground state from the old to current DMEV basis, 
followed by the evaluation of $E_{gs}$ as, 
$E_{gs} ~=~ \frac{\bra \psi_{gs} \mid H \mid \psi_{gs} \ket}{\bra \psi_{gs} \mid \psi_{gs} \ket}$;
$H$ being the super-block Hamiltonian in the current DMEV basis.

As the number of target states used for constructing the reduced block density matrices 
increases, the overall computational time for the infinite and finite DMRG schemes also increases.
The infinite part of DTWT algorithm uses three target states while the finite part needs 2{\it p} states
as target states. Hence, the finite part of our algorithm is slower compared to the infinite part. To
overcome this problem, we have used the {\it window modification} that we developed and employed in the
context of LXW algorithm. Instead of retaining all the 2{\it p} time-dependent states as target states, we
keep $\frac{2p}{10}$ target states; instead of 10, other variations can also be employed. Incorporation
of the {\it window modification} into the finite part of our algorithm has an interesting consequence, 
namely, the 2{\it p} $< ~m$ condition can be replaced by $\frac{2p}{10} ~<~ m$, thereby making it feasible to
increase the length of both the double (2$\Delta t$) and single ($\Delta t$) time windows. 

The time step error associated with the fourth-order R-K technique is $O(\Delta t^{5})$ while that associated
with the MSD2 scheme is $O(\Delta t^{3})$. Thus, the former time propagation scheme is more accurate than 
the latter. However, the TDSE solver required for constructing the Hilbert space of the time-evolving wave 
function need not be very accurate compared to TDSE solver needed to propagate the wave packet in time. 
Hence we have used two different time propagation techniques for the two different windows present in our 
algorithm. For the double time window (2$\Delta t$) evolution which is used for basis adaptation, we have 
used the second-order multistep differencing (MSD2) scheme \cite{cakmak}, given by
\beq
\mid \psi(t+\Delta \tau) \ket ~=~ -2iH\Delta \tau \mid \psi(t) \ket + \mid \psi(t-\Delta \tau) \ket,
\eeq 
which uses one sparse matrix-vector multiplication (SMVM) operation for every propagated time step 
$\Delta \tau$. The two states $\mid \psi(t) \ket$ and $\mid \psi(t-\Delta \tau) \ket$ are obtained once in 
the beginning of every time evolution using fourth-order R-K technique. Hence, every time evolution from 
{\it t} $\rightarrow$ ($t$+2$p\Delta \tau$) involves $(8+[2p-2]) ~=~ (2p-6)$ SMVMs. For the final single time
window propagation we employ the fourth-order R-K technique as the TDSE solver. Hence, every time evolution 
from $t ~\rightarrow$ ($t+p\Delta \tau$) would require $4p$ SMVMs. Therefore, each single time window 
propagation using the MSD2 and R-K techniques involves 
$\biggl([2p-6] \times 4[\frac{N_{s}}{2}-2] + 4p \biggr)$ SMVMs. 

\section{COMPARISON OF THE DTWT SCHEME WITH THE LXW AND TST TECHNIQUES}

In this section we present a comparative analysis of the computational efficiency of our DTWT scheme with the
LXW and TST schemes. In the TST technique, as pointed by White, the wave function is propagated in time after
either one or several half-sweeps depending on the how many half sweeps are needed to update the Hilbert 
space of the wave function adequately. Each half sweep (according to our definition of half sweep) needs 
($\frac{N_{s}}{2}-2$) finite DMRG steps. If the wave function is propagated in time by $\Delta \tau$ after 
every half sweep, then for the total time propagation, $Np$($\frac{N_{s}}{2}-2$) finite DMRG steps are 
needed. However, if the wave function is propagated after every $q$ half sweeps, then 
$Npq$($\frac{N_{s}}{2}-2$) finite DMRG steps are needed for propagating the initial wave packet over the 
entire time interval, 0 to T. In the DTWT technique, for every two full sweeps ($q$ = 4) the wave function is
evolved by a time interval $\Delta t$ which implies that $Nq$($\frac{N_{s}}{2}-2$) DMRG steps are needed for
evolving the initial state over the entire time interval. Therefore, the ratio of DMRG steps in DTWT to TST 
is either $\frac{4}{pq}$, if $q ~\ne 1$, or ($4/p$), if $q ~=~ 1$ in the TST procedure. Since $p$ is chosen 
to be $\ge ~10^{2}$, the DTWT scheme is more efficient than the TST technique by a factor of $\ge$ 25. The 
actual CPU time involved in the time evolution by the fourth-order R-K procedure is the same in both TST and 
DTWT schemes. In case of the LXW algorithm, the DMRG steps involved are fewer for a system size $N_{s}$, the 
number of DMRG steps is only $(\frac{N_{s}}{2}-1)$. However, we evolve the wave packet at each system size 
and this is the CPU intensive part of the calculation. Besides, the target states in LXW scheme is equal to 
number of time intervals $N$ (specially for systems with long-range interactions), which is $\sim 10^{4}$ or 
more. Therefore the DMEV cut-off required for comparable accuracy is huge $(\sim 10^{4})$ and unattainable. 
This leads to large errors in the LXW scheme at long times.
To test the method we compared the DTWT results for various {\it m} values with exact time evolution of a 
$14$-site H\"uckel as well as PPP chain. We also compared these results with those obtained by the TST method
for a $14$-site H\"uckel chain and LXW for a $40$-site Hubbard chain with $\frac{U}{\mid t \mid}$ = 4. 
To test whether our DTWT scheme is applicable to other geometries, we have compared the exact time evolution 
of a $12$-site biphenyl molecule and a $14$-site stilbene molecule (see Fig. 4), both modeled by the PPP 
Hamiltonian, with different DMEV cut-offs, {\it m}. We have dealt with H\"uckel chains of length up to 
$14$-sites since the number of states (although known exactly) become too large and the computations become 
cumbersome for time propagation of the wave packet.

The model Hamiltonians used in this study are the tight-binding Hamiltonian \cite{huc1,huc2}, also known as 
the H\"uckel Hamiltonian to chemists, the Hubbard \cite{hub1,hub2,hub3} and PPP Hamiltonian \cite{ppp1,ppp2}.
The second quantized Hamiltonians for these models are give below \cite{surjan}:
\begin{gather}
H_{\text{H\"uckel}} ~=~ \sum_{<ij>,\sigma=\uparrow,\downarrow} t_{ij} (c^{\dagger}_{i,\sigma}c_{j,\sigma} ~+~ c^{\dagger}_{j,\sigma}c_{i,\sigma}), \nonumber \\
H_{\text{Hubbard}} ~=~ H_{\text{H\"uckel}} ~+~ \sum_{i} U_{i} n_{i,\uparrow}n_{i,\downarrow}, \nonumber \\
H_{\text{PPP}} ~=~ H_{\text{Hubbard}} ~+~ \sum_{i>j} V_{ij} (n_{i}-z_{i})(n_{j}-z_{j}). 
\end{gather} 
In the first equation, the summation is restricted to bonded neighbors $<ij>$, $t_{ij}$ is the hopping 
integral between bonded neighbors. $U_{i}$ is the on-site Coulomb repulsion term (Hubbard $U$-term) of 
the $i$th site, $c^{\dagger}_{i,\sigma}$ ($c_{i,\sigma}$) creates (annihilates) an electron with spin 
$\sigma$ at (from) the $i$th site, $n_{i\sigma}$ are the corresponding number operators, $n_{i} = 
(n_{i,\sigma} ~+~ n_{i,-\sigma})$ is the charge density at site $i$, $V_{ij}$ is the intersite Coulomb 
repulsion between lattice sites ($i, j$), and $z_{i}$ is the on-site chemical potential of site $i$. For 
homogeneous systems, $U_{i}$ = $U$, and is a measure of the Coulomb repulsion between two electrons of 
opposite spins occupying the same site. ($U/t$) characterizes the strength of electron correlations. In case 
of homogeneous $sp^{2}$ carbon systems, all sites are singly occupied for charge neutrality and hence 
$z_{i}$ = 1, $\forall ~i$. The inter-site electron-electron repulsion term ($V_{ij}$) in the PPP model is 
phenomenologically interpolated between $U$ for zero separation and $\frac{e^2}{r_{ij}}$ for inter-site 
separation $r_{ij} ~\rightarrow ~\infty$; thus, the explicit evaluation of the repulsion integrals is 
avoided. There are two widely used interpolation schemes for evaluating $V_{ij}$: Ohno scheme \cite{ohno} 
and, the Mataga-Nishimoto scheme \cite{mataga}. In the Ohno interpolation scheme which we use, the $V_{ij}$ 
term is given by
\beq
V_{ij} ~=~ 14.397 \biggl[ \biggl(\frac{28.794}{U_{i}+U_{j}} \biggr)^{2} ~+~ r^{2}_{ij} \biggr]^{-1/2}
\eeq 
{\noindent and decays more rapidly than the Mataga-Nishimoto formula which is shown below.}
\beq
V_{ij} ~=~ \biggl[ \frac{2.0}{U_{i}+U_{j}} ~+~ \frac{r_{ij}}{14.397} \biggr]^{-1}.
\eeq  
In both the above interpolation formulas, $r_{ij}$ is measured in angstrom ($\AA$) while $U$ and $V_{ij}$ are
measured in electron volt (eV). In the Hubbard model calculations, we have used $\frac{U}{\mid t \mid}$ = 4 
while in the PPP model we have used standard parameters ($U$ = 11.26 eV, $t$ = -2.4 eV, and $r$ = 1.397 
$\AA$; $r$ being the uniform C-C bond length) and Ohno parametrization.
\begin{figure}
\epsfig{file=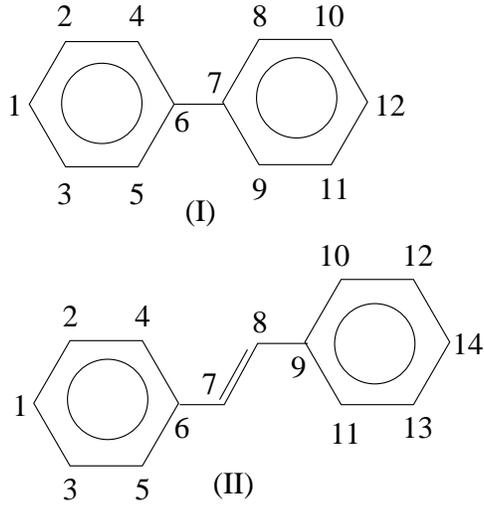, width = 2.5 in}
\caption{Molecular structures of (i) biphenyl and (ii) stilbene molecules. Biphenyl and stilbene are the 
monomers of polyparaphenylene and polyparaphenylvinylene polymers, respectively. Site numbering scheme as 
employed by us, is also shown in the figure.}
\label{figure:4}
\end{figure}
In Fig. 5 we compare the DTWT and TST schemes for different DMEV cut-offs {\it m} for a H\"uckel chain
of $14$ sites, with exact results. From Fig. 5 it is clear that for a given system size and {\it m}, the 
DTWT algorithm has a better accuracy than the TST algorithm. The results for the Hubbard chain (Fig. 6) of 
$40$ sites show a smooth convergence in the DTWT scheme as {\it m} is increased. This gives confidence in 
the DTWT scheme. The LXW method with {\it m} = 200 differs from our results quantitatively at long times 
(Fig. 6). Our results for the $14$-site PPP chain indicate that long-range interacting models require higher 
{\it m} for attaining the same convergence as with the nearest-neighbor models (Fig. 7). In Fig. 8  we show 
the time evolution of the charge and spin densities at the sites numbered according to Fig. 4. Fig. 8 also 
shows that molecular topologies which are not linear, require larger DMRG basis dimension to attain 
accuracies comparable to open chain system.
\begin{figure}
\begin{center}
\vspace*{-1.7cm}
\epsfig{file=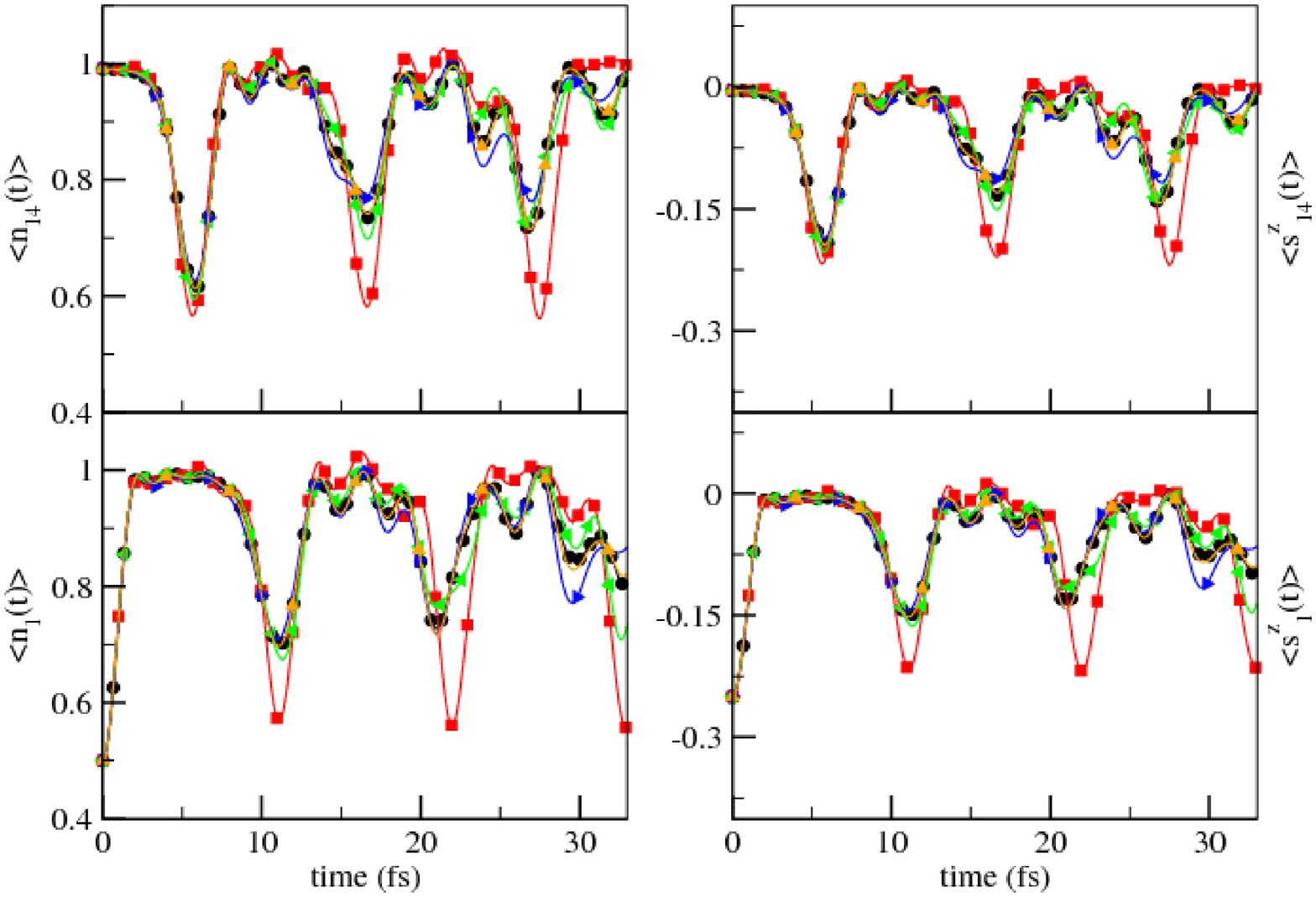, width = 4.5 in} \\
\caption{(Color online)
Time evolution profiles of charge (left curve) and spin (right curve) densities at first and last sites of a 
$14$-site H\"uckel chain, using three different techniques. The color codings for the curves are as follows: 
black curve with circles = exact; red curve with squares=TST(m=64); green curve with left triangles=TST(m=100); blue curve with right triangles=DTWT(m=64); orange curve with up triangles=DTWT (m = 100). The m=100 DTWT 
curve is almost indistinguishable from the exact curve. $\Delta t$ = 0.066fs and $\Delta \tau$ = $\frac{\Delta t}{10}$ are used in the TST technique. In the DTWT scheme, $\Delta \tau$ = 0.066fs is employed.}
\label{figure:5}
\end{center}
\vspace*{0.3cm}
\begin{center}
\epsfig{file=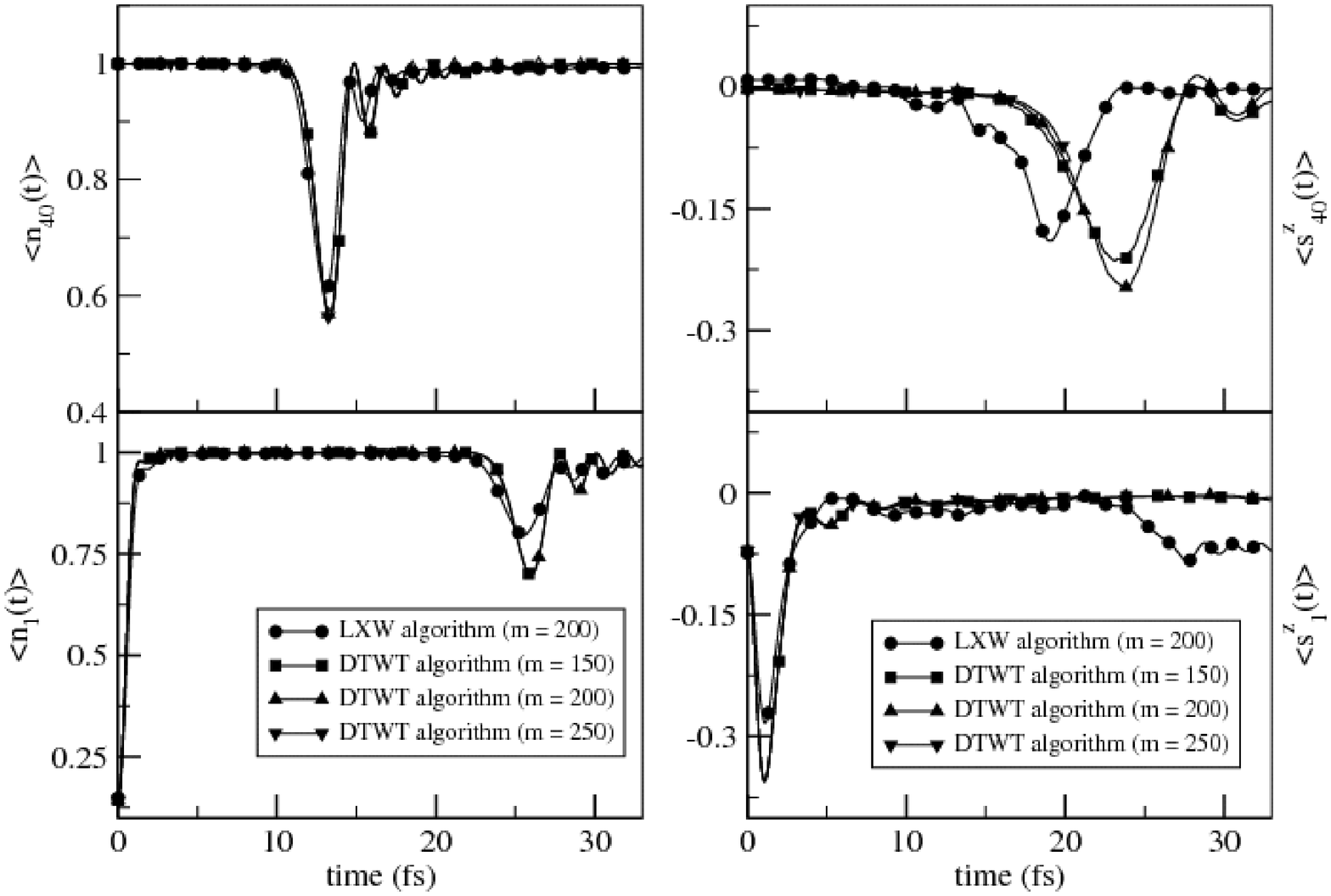, width = 4.5 in}
{\caption{\small 
Comparison of time evolution of charge (left curve) and spin (right) densities at the first and last 
sites of a 40-site Hubbard chain, using the LXW and DTWT techniques. In both the DTWT and LXW schemes, 
$\Delta \tau$ is taken to be 0.066fs.}}
\label{figure:6}
\end{center}
\end{figure}

\begin{figure}
\begin{center}
\vspace*{-2.0cm}
\epsfig{file=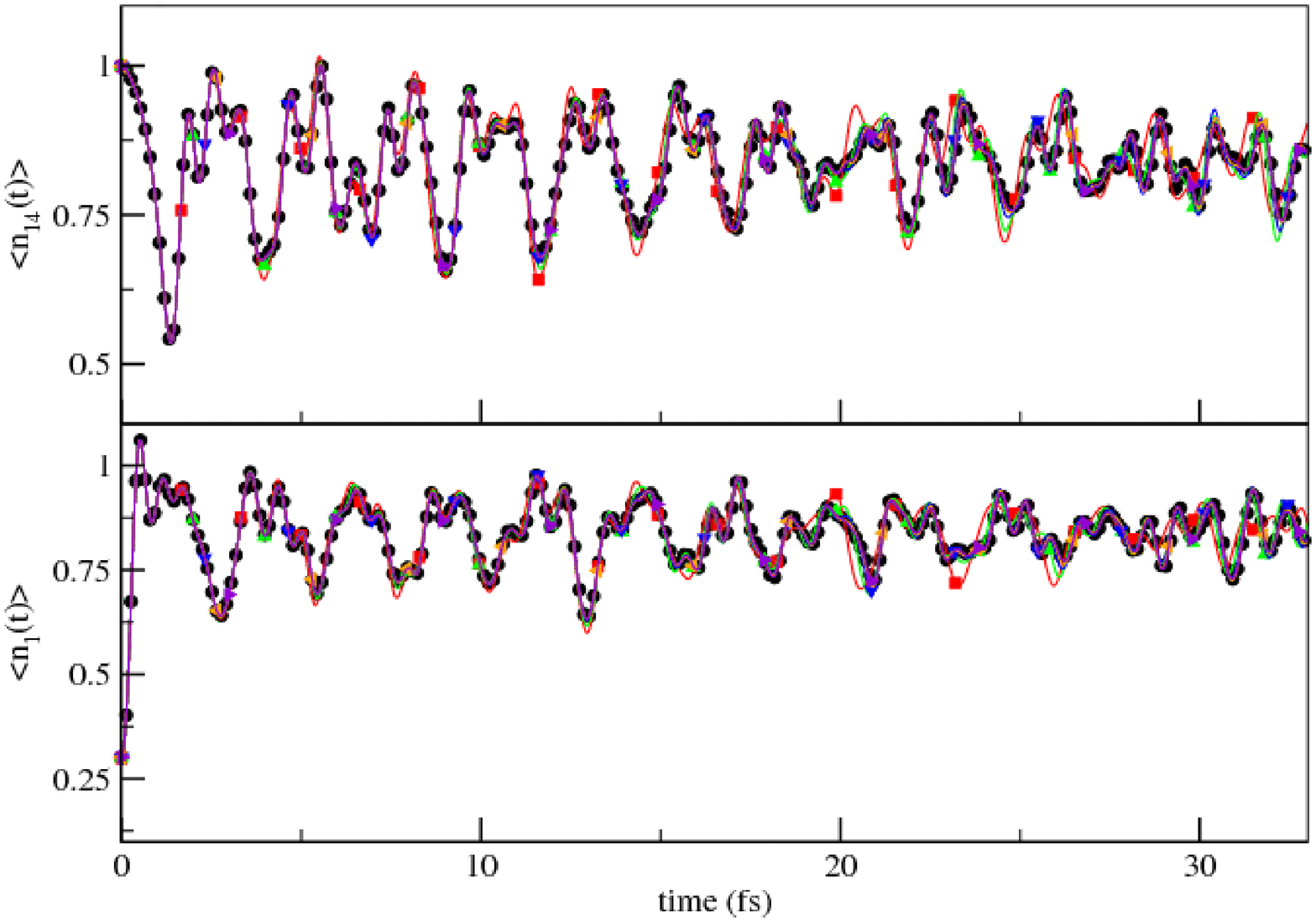, width = 5.0 in} \\
\vspace*{1.0cm}
\epsfig{file=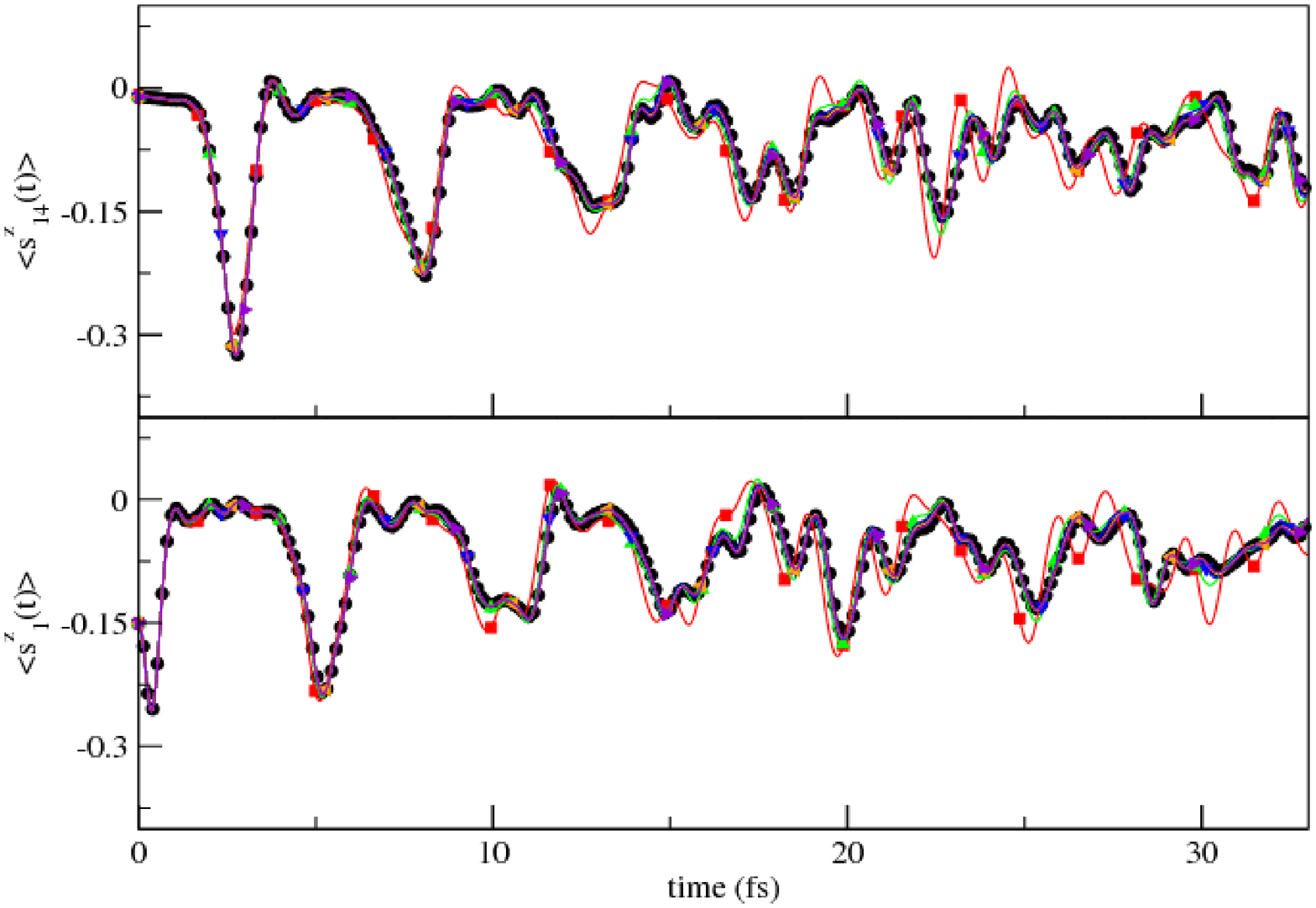, width = 5.0 in}
\caption{(Color online)
Comparison of exact versus DTWT time evolution of charge (top curve) and spin (bottom curve) densities at 
first and last sites of a $14$-site PPP chain. $\Delta \tau$ is chosen to be 0.0066fs. The color coding is as
follows: black curve with circles = exact; red curve with squares=DTWT(m=64); green curve with up triangles=DTWT(m=100); blue curve with down triangles=DTWT(m=150); orange curve with left triangles=DTWT(m=200); violet 
curve with right-triangles=DTWT(m=250). It is observed that for {\it m} $\ge$ 100, the curves converge 
towards the exact time evolution, and  for {\it m}=200, the DTWT curve is coincident with the exact curve.}
\label{figure:7}
\end{center}
\end{figure}

\begin{figure}
\begin{center}
\vspace*{-1.8cm}
\epsfig{file=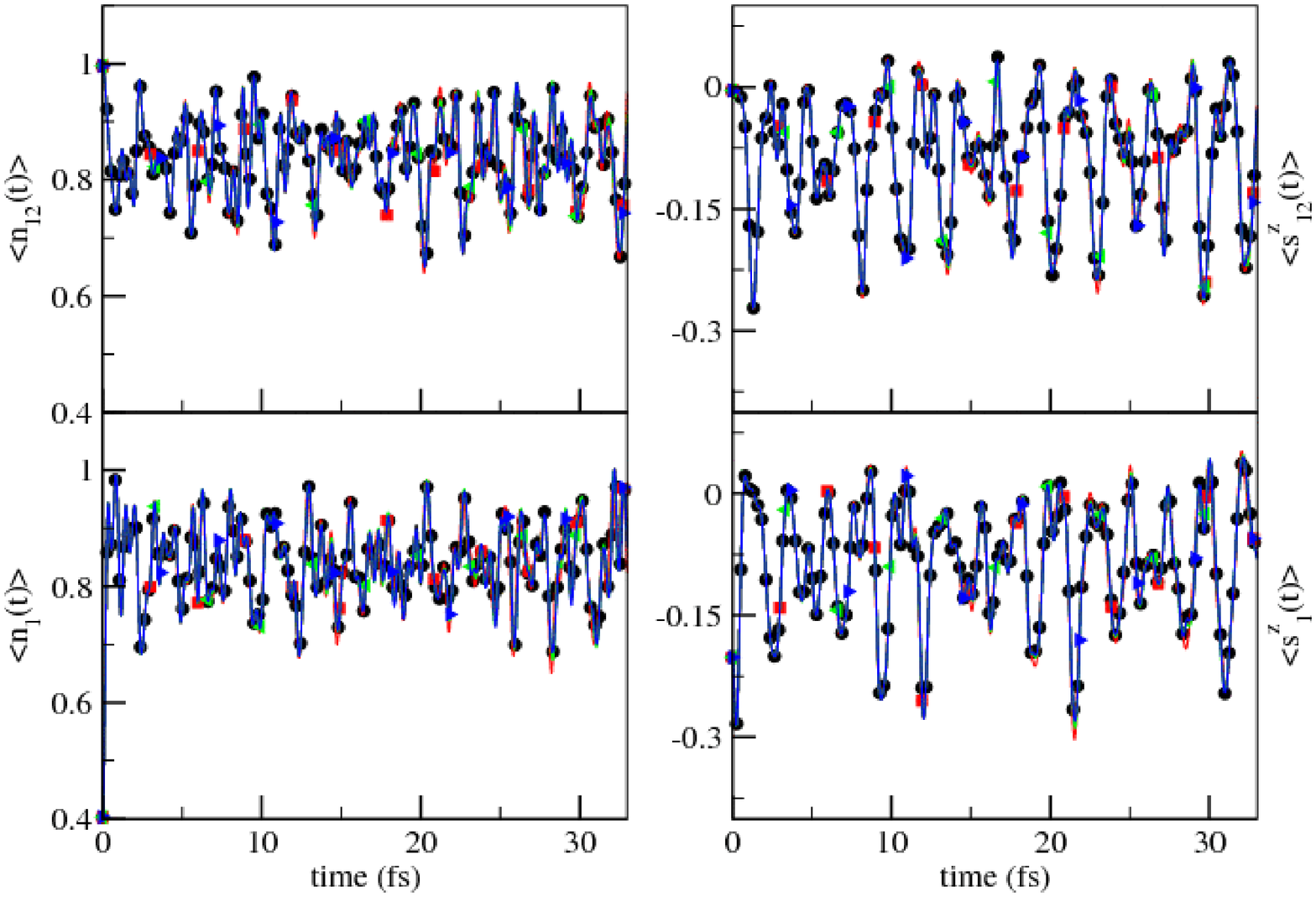, width = 5.0 in} \\
\vspace*{1.0cm}
\epsfig{file=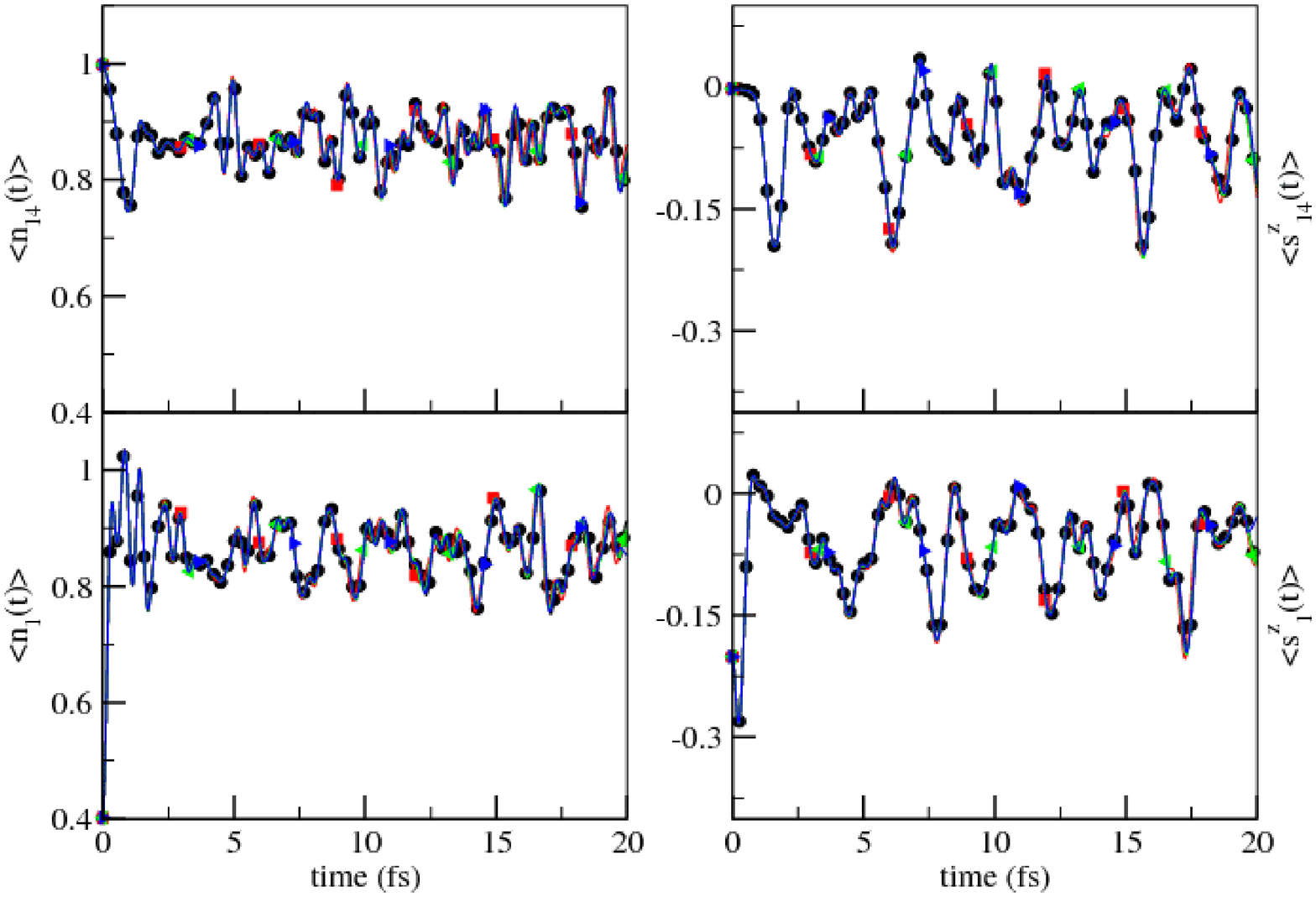, width = 5.0 in}
\caption{(Color online)
Comparison of exact versus DTWT time evolution of charge (left curve) and spin (right curve) densities at 
first site and last site of biphenyl system (top) and stilbene molecule (bottom) whose structures are given 
in Fig. 4. The color coding is as follows: black curve with circles: exact; red curve with squares=DTWT(m=150); green curve with left triangles=DTWT(m=200); blue curve with right triangles=DTWT(m=250). }
\label{figure:8}
\end{center}
\end{figure}

\newpage
\section{REAL-TIME DYNAMICS OF SPIN-CHARGE SEPARATION IN PPP CHAINS}

The quantities we have studied using our time-dependent DMRG algorithm are the time dependence of the site 
charge and site spin densities. The charge and spin densities at the $i$th site of a system at time $t$ are,
respectively,
\beq
\bra n_{i}(t) \ket ~=~ \bra \psi(t) \mid (n_{i \sigma} + n_{i -\sigma}) \mid \psi(t) \ket 
\eeq
{\noindent and}
\beq
\bra s^{z}_{i}(t) \ket ~=~ \bra \psi(t) \mid (n_{i \sigma} - n_{i -\sigma}) \mid \psi(t) \ket, 
\eeq 
where $n_{i \sigma}$'s are the number operators with spin $\sigma$ and $\mid \psi(t) \ket$ is the wave packet 
evolved in time. Although we have calculated charge (spin) density at all sites of the systems studied in 
this work, we focus on $\bra n_{1}(t) \ket$, $\bra n_{L}(t) \ket$, $\bra s^{z}_{1}(t) \ket$, and
$\bra s^{z}_{L}(t) \ket$, which suffice to investigate spin-charge separation in the PPP model; $1$ and $L$ 
correspond to the two terminal sites of the chain. We have done two sets of calculations for each of the 
systems mentioned above, keeping the DMEV basis sizes at 200 and 250, and we find the results converge. Here,
we present data obtained with the smaller DMEV basis, namely, with 200 states.

Consider the ground state of a half-filled system with particle density, $n ~=~ (\frac{N_{e}}{2 N_{s}}) ~=~ 
1/2$, $N_{e}$ and $N_{s}$ are, respectively, the total number of electrons and sites in the system. The 
ground state of this neutral system being an eigenstate of the Hamiltonian, is a stationary state. When an 
electron with a definite spin is either added or removed at a site, from the ground state of the system, a 
new state forms which is no longer an eigenstate of the  governing Hamiltonian. This is a transient 
(non-stationary) state, popularly known as ``wave packet'' in the literature, whose time evolution we study 
for a finite time. In our studies, we annihilate an up-spin electron from site $1$ of the systems we have 
considered and this corresponds to injecting a downspin hole into the system at site $1$. In the singlet 
($\bra S^{z}_{tot} \ket$ = 0) ground state of the neutral system, $\bra n_{i} \ket$ $\approx$ 1.0 and 
$\bra s^{z}_{i} \ket$ $\approx$ 0.0, $\forall ~i$ $\in$ $[$1,$N_{s}]$, $N_{s}$ = total number of sites in the
system. The injected hole at $t$ = 0 is localized at the injection site (site 1). In the initial state 
(wave packet), except at the injection site, all other sites have charge (spin) density of 1.0 (0.0). 
Evolution of this non-stationary state in time, leads to change in the charge (spin) density distribution of 
the system, and temporal variation in these dynamical quantities is viewed as propagation of the injected 
hole from site $1$ to site $L$. Since our systems are not connected to reservoirs, the particle number is 
fixed and the hole is reflected from the two ends of the system. Hence, time evolution profiles of charge
(spin) density consists of a series of small and large maxima and minima. All the systems studied by us are 
homogeneous, bipartite lattices, implying that all sites are equivalent and the lattice possesses 
electron-hole symmetry at half filling.

In the absence of electron-electron correlation, temporal variation in charge and spin densities of the 
injected hole are identical. This is because the charge and spin degrees of freedom of the hole propagate 
with the same velocity, namely, the Fermi velocity ($\vartheta_{F}$). Electron-electron correlation decouples
these intrinsic degrees of freedom into two separate elementary excitations: holon (carrying charge but no 
spin) and spinon (carrying spin but no charge). This decoupling is known as spin-charge separation and has 
been widely studied in the literature. Thus, in the presence of electron-electron correlation, the time 
evolution profiles of charge and spin densities of the system are different from each other and is recognized
as a manifestation of spin-charge decoupling. In order to address the issue of spin-charge separation in the 
PPP model for a given topology, we have focused on two major extremal points in the time evolution profiles 
of charge and spin densities of the injected hole. These give us an estimate of the velocity of the charge 
and spin of the hole, and are therefore helpful in analyzing the spin-charge separation phenomena in the PPP 
model. These points correspond to the first major minima (dip) in the time evolution profiles of 
$\bra n_{1}(t) \ket$ and $\bra s^{z}_{1}(t) \ket$, and $\bra n_{L}(t) \ket$ and $\bra s^{z}_{L}(t) \ket$, 
respectively, at $t ~\ne$ 0. The time taken for the $t ~\ne$ 0 first minima to appear in the time evolution 
profiles of $\bra n_{1}(t) \ket$ and $\bra s^{z}_{1}(t) \ket$ is associated with the event of the charge and
spin degree of freedom of the hole propagating from site $1$ to site $L$ and returning to site $1$, 
respectively. The time taken for this dip to appear in the charge (spin) density profile is designated as 
$\tau^{h}_{2L}$ ($\tau^{s}_{2L}$). The charge velocity deduced from this time is 
$\vartheta^{h}_{2L} ~=~ (\frac{2L}{\tau^{h}_{2L}})$ while the spin velocity is 
$\vartheta^{s}_{2L} ~=~ (\frac{2L}{\tau^{s}_{2L}})$. The first minima in the temporal profiles of 
$\bra n_{L}(t) \ket$ ($\bra s^{z}_{L}(t) \ket$) appears when the charge (spin) of the injected hole migrates 
from site 1 to site $L$ of the system. The time taken for this event is denoted as 
$\tau^{h}_{L}$ ($\tau^{s}_{L}$); charge and spin velocities associated with this event are,
$\vartheta^{h}_{L} ~=~ (\frac{L}{\tau^{h}_{L}})$ and $\vartheta^{s}_{L} ~=~ (\frac{L}{\tau^{s}_{L}})$,
respectively. 

Spin-charge separation in correlated one-dimensional systems has been studied using the DMRG technique
by Kollath {\it et al.} \cite{kollath} and Ulbricht and Schmitteckert \cite{peter}. Kollath {\it et al.}
studied the dynamics of a wave packet obtained by introducing a particle at $t$ = 0 in the middle of a
long Hubbard chain equilibrated in a spin-dependent site energy acting on the system at time t ~$<$ 0
and turned off at $t$ = 0. Both for different fillings and different chain lengths, they observed that the
charge and spin velocities were different. Ulbricht and Schmitteckert observed spin-charge separation
in a transport simulation involving non-interacting leads and interacting system. The initial wave packet
consists of a particle with Gaussian probability distribution, moving towards the interacting region,
added to one of the leads. The time-dependent study of site spin and charge densities show a separation of 
spin and charge. Our studies are carried out on molecular systems with long-range interactions and the 
charge injection, is made at the end of the chain.

For studying dynamics in the PPP model, we have considered regular polyene chains (uniform transfer integral)
. This implies that all bond lengths are equal. Although polyenes are experimentally known to exist in 
dimerized form, yet our focus being on the algorithm, we have not considered dimerized chains in the present 
study. Analytical expressions for the velocity of the charge degree of freedom (holon), $\vartheta_{h}$, and 
spin degree of freedom (spinon), $\vartheta_{s}$, in the large $U$ limit of the one-dimensional Hubbard model 
exist in literature \cite{jagla,coll},
\beq
\vartheta_{h} = 2 \mid t \mid \sin(\pi n); ~~~~\vartheta_{s} = \frac{2 \pi \mid t \mid^{2}}{U} \biggl[1 - \frac{\sin(2 \pi n)}{2n} \biggr],  
\eeq
where $t$ and $U$ are, respectively, the nearest-neighbor hopping matrix element and the on-site Coulomb 
repulsion term, and $n$ is the particle density ($n ~\leq ~1$). From the above expressions it is evinced that
while $\vartheta_{h} \propto$ $\mid t \mid$, $\vartheta_{s} \propto \frac{\mid t\mid^{2} }{U}$. Thus, for a 
given value of $t$, increasing $U$ is tantamount to decreasing the velocity of the spinon. Furthermore, as 
$U ~\rightarrow \infty$, $\vartheta_{s} \rightarrow$ 0 and we reach the atomic limit. However, analysis of 
the charge ($\vartheta_{h}$) and spin ($\vartheta_{h}$) velocities of the injected hole for the 
one-dimensional PPP model do not exist. The PPP model is basically considered as a Hubbard model augmented 
with an additional long-range Coulomb repulsion term, which leads to a renormalized $U$ in the Hubbard model.
However, the physics of the PPP and Hubbard models are quite different; the potential energy of a 
configuration not only depends on the number of doubly occupied sites but also on the actual distribution of 
these sites. Thus, it is not possible to naively map the PPP models to effective Hubbard models. This makes a
comparison of the charge and spin velocities of PPP model with those of the Hubbard model interesting. We 
have carried out this comparison of the PPP results with those of the Hubbard model with 
$\frac{U}{\mid t \mid}$ = 2.0, 4.0, and 6.0, which we have published previously \cite{dutta}. 
\begin{figure}
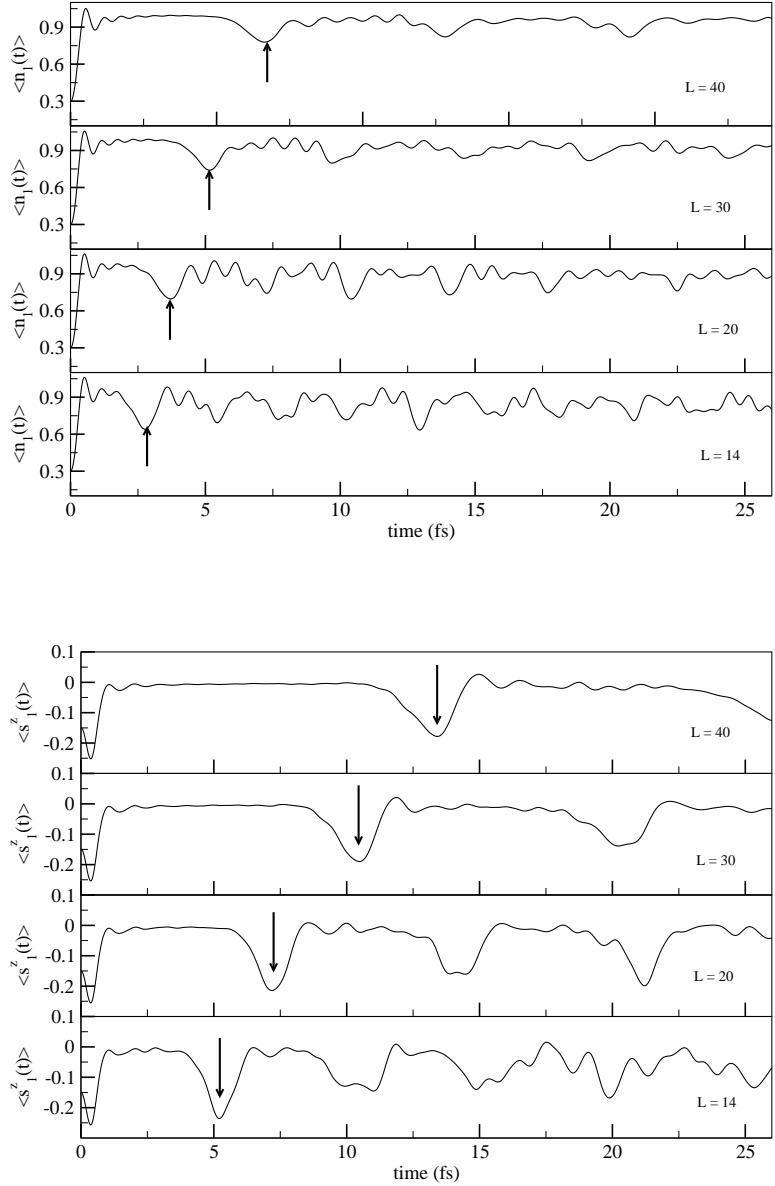

\begin{center}
\vspace*{-2.0cm}
\epsfig{file=chrgden_site1_m200.eps, width = 4.0 in} \\
\vspace*{1.3cm}\epsfig{file=spinden_site1_m200.eps, width = 4.0 in}
\caption{
Time evolution profiles of $\bra n_{1}(t) \ket$ (left curve) and $\bra s^{z}_{1}(t) \ket$
(right curve) for regular PPP chains of length 14, 20, 30, and 40. The position of the $\tau^{h}_{2L}$
and $\tau^{s}_{2L}$ are indicated with arrow.}
\label{figure:9}
\end{center}
\end{figure}

\begin{figure}
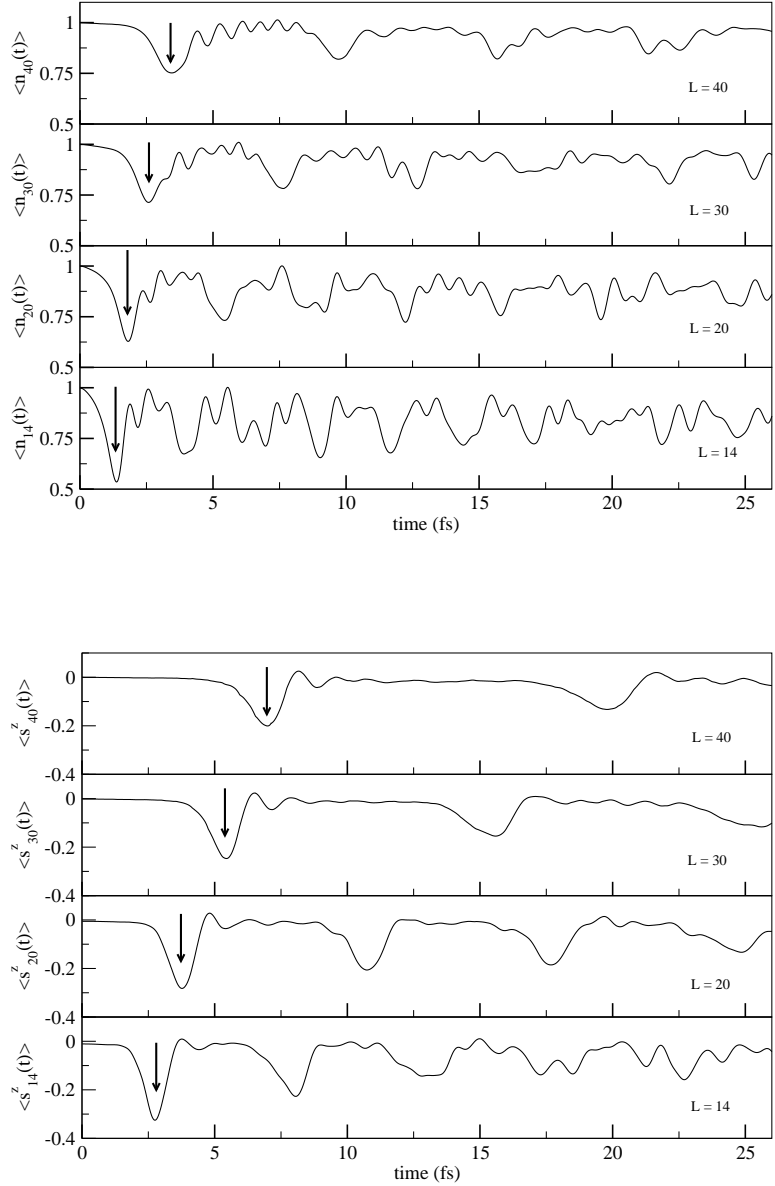

\begin{center}
\vspace*{-1.8cm}
\epsfig{file=chrgden_siteN_m200.eps, width = 4.0 in} \\
\vspace*{1.5cm}\epsfig{file=spinden_siteN_m200.eps, width = 4.0 in}
\caption{
Time evolution profiles of $\bra n_{L}(t) \ket$ (left curve) and $\bra s^{z}_{L}(t) \ket$
(right curve) for regular PPP chains (L = 14,20,30,40). The position of the $\tau^{h}_{L}$
and $\tau^{s}_{L}$ are indicated with arrow. }
\label{figure:10}
\end{center}
\end{figure}

In Figs. 9 and 10 we show the time evolution of $\bra n_{1}(t) \ket$, $\bra s^{z}_{1}(t) \ket$ and
$\bra n_{L}(t) \ket$, $\bra s^{z}_{L}(t) \ket$ for different chain lengths of the PPP model. Time taken for 
the second minimum to appear in the temporal profiles of $\bra n_{1}(t) \ket$ and $\bra s^{z}_{1}(t) \ket$ 
(Fig. 9) gives $\tau^{h}_{2L}$ and $\tau^{s}_{2L}$, respectively, and are the time taken by the charge and 
spin to travel from site $1$ to site $L$ and back to site $1$. Two observations can be made from the results:
$\tau^{h}_{2L} > \tau^{s}_{2L}$ for all the chains considered, that is, the hole charge moves faster than the
hole spin in the PPP model also, just as in the case of the Hubbard model and the velocities 
$\vartheta^{h}_{2L}$ and $\vartheta^{s}_{2L}$ are weakly dependent on the system size. Similarly, time taken 
for appearance of the first minimum in the time evolution profiles of $\bra n_{L}(t) \ket$ and 
$\bra s^{z}_{L}(t) \ket$ (Fig. 10) provides $\tau^{h}_{L}$ and $\tau^{s}_{L}$ respectively. The charge and 
spin velocities of the hole, $\vartheta^{h}_{L}$ and $\vartheta^{s}_{L}$, calculated from the dynamics of the
$L$th site as $\biggl(\frac{L}{\tau_{L}^{h}}\biggr)$ and $\biggl(\frac{L}{\tau_{L}^{s}}\biggr)$, 
respectively, agree with the velocities calculated from the $\tau^{h}_{2L}$ and $\tau^{s}_{2L}$ values 
($\vartheta^{h}_{2L}$ and $\vartheta^{s}_{2L}$), corresponding to the dynamics of the first site. We also 
note that finite-size effects are weak from the linear dependence of $\tau^{h/s}_{L/2L}$ on the system size 
and nearly system size independent ratio of $\vartheta^{h/s}_{L/2L}$ (Fig. 11). For the standard PPP 
parameters, the charge velocity is nearly twice the spin velocity (see Table. I). To further investigate the 
finite-size effects on correlation, we plotted the ratio of hole (spin) velocity for the 40-site chain to 
that of 20-site chain $\biggl(\frac{\vartheta^{h/s}_{40}}{\vartheta^{h/s}_{20}}\biggr)$ versus 
$\frac{U}{\mid t \mid}$, (Fig. 12). In the case of holes it is found that this ratio is maximum 
for $\frac{U}{\mid t \mid}$ = 0.0 and significantly decreases as $\frac{U}{\mid t \mid}$ increases
and beyond $\frac{U}{\mid t \mid}$ = 4.0 it is nearly constant, with a value close to 1.0 signifying
infinite chain behavior. However, the spin velocity ratio shows a dip at $\frac{U}{\mid t \mid}$ = 4.0
and increases significantly as $\frac{U}{\mid t \mid}$ increases. Thus, it appears that the spin properties
show stronger finite-size effects than charge properties. In case of the PPP model, it is observed that the 
ratio of $\biggl(\frac{\vartheta^{h}_{40}}{\vartheta^{h}_{20}}\biggr)$ $\approx$ 
$\biggl(\frac{\vartheta^{s}_{40}}{\vartheta^{s}_{20}}\biggr)$ $\approx$ 1.1, and the finite-size effects are
weak. The finite-size effect in the H\"uckel model arises from the kinetic-energy term, longer the chain,
lower the kinetic energy due to delocalization, and the system is stable. In the Hubbard model, the 
finite-size effects arise due to suppression of charge fluctuations which is more effective in longer chains
due to delocalization. However, in PPP models, the charge fluctuations are better accommodated due to 
long-range interactions and the kinetic term is not as dominant as non-interacting models. Therefore,
we can anticipate weaker finite-size effects in the PPP model than either the H\"uckel or Hubbard model.

In the extreme case of correlation strength in the Hubbard model $\frac{U}{\mid t \mid} ~\rightarrow 
~\infty$, the charge velocity will be a constant near $n ~\approx$ 0.5 and the spin velocity tends to zero 
[Eq. (19)]. The PPP model therefore appears to be in the intermediate correlation regime. In order to compare
the PPP model with the Hubbard model, we have also carried out the time dependent study for different 
$\frac{U}{\mid t \mid}$ values for the Hubbard model. In Fig. 13 we compare the time evolution of 
$\bra n_{L}(t) \ket$ and $\bra s^{z}_{L}(t) \ket$ of the Hubbard model with $\frac{U}{\mid t \mid}$ = 2.0, 
4.0, and 6.0, and the PPP model for different chain lengths. We find that for all the three correlation 
strengths of the Hubbard model considered, the charge as well as spin, move slower than in the PPP model. The
ratio of the charge velocity to the spin velocity for the Hubbard model is presented in Table I. We see that 
for the Hubbard model with strong correlation strength, $\frac{U}{\mid t \mid}$ = 6.0, this ratio is closer
to that of the PPP model. It is usually the practice to compute the effective $\frac{U}{\mid t \mid}$ in an
extended range model to be $\frac{(U-V_{1,2})}{\mid t \mid}$, where $V_{1,2}$ is the first neighbor 
electron-electron interaction range. This simplistic interpretation of long-range correlations only 
renormalizing the $U$ value of the Hubbard model seems to be erroneous in treating the dynamics of doped 
holes. We note that very strong on-site correlation of $\frac{U}{\mid t \mid} ~>~ $ 6.0 is required to 
reproduce the ratio of the charge and spin velocities in the PPP model. However, U$_{eff}$  computed as 
$\frac{(U-V_{1,2})}{\mid t \mid}$ for standard parameters is 1.52, underestimates severely the strength of
correlations. It appears that the dependence of correlation on energy of the actual charge distribution 
enhances the role of correlations in the dynamics of charge and spin transport.  

\begin{figure}
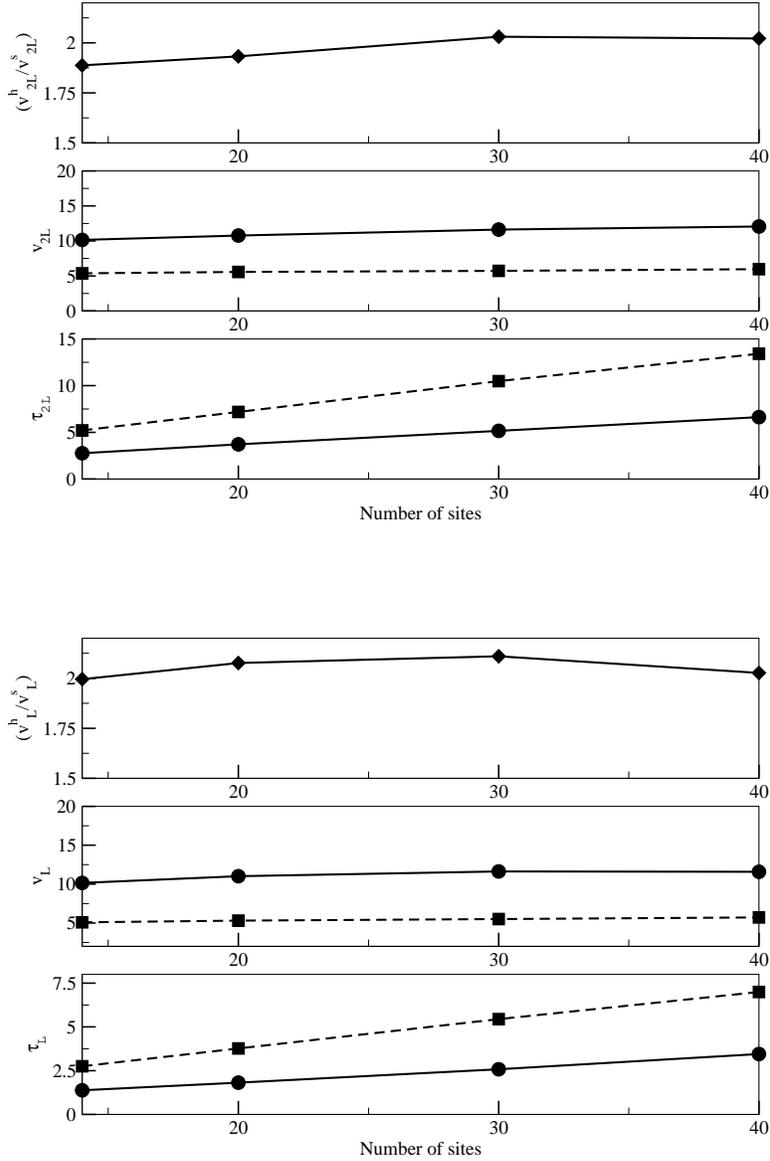

\begin{center}
\vspace*{-2.5cm}
\epsfig{file=velocityntime_vs_2L.eps, width = 4.0 in} \\
\vspace*{1.5cm}\epsfig{file=velocityntime_vs_L.eps, width = 4.0 in}
\caption{
Variation in $\tau^{h/s}_{L/2L}$, $\vartheta^{h/s}_{L/2L}$ and  the ratio 
$\biggl(\vartheta^{h}_{L/2L}/\vartheta^{s}_{L/2L}\biggr)$ with chain length in the PPP model. Solid curves
with circles represent $\vartheta^{h}_{2L}$ (top) and $\vartheta^{h}_{L}$ (bottom). Dashed curves with 
squares represent $\vartheta^{s}_{2L}$ (top) and $\vartheta^{s}_{L}$ (bottom).}
\label{figure:11}
\end{center}
\end{figure}

\begin{figure}
\begin{center}
\epsfig{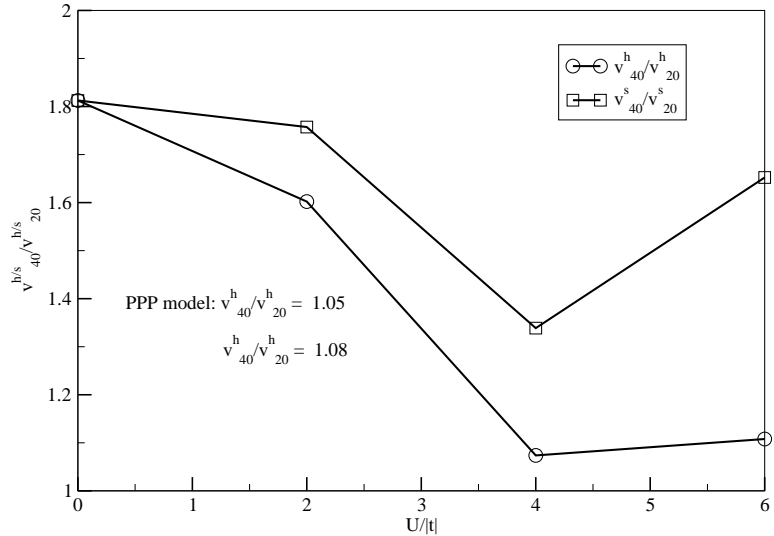}
\caption{
Finite size effect of hole and spin velocities as a function of correlation strength $\frac{U}{\mid t \mid}$.
Ratio of the hole (spin) velocity in a $40$-site chain to those in a $20$-site chain, 
$\biggl(\frac{\vartheta^{h/s}_{40}}{\vartheta^{h/s}_{20}}\biggr)$, is plotted on the Y-axis. The ratios for 
the PPP model are quoted in the figure.} 
\label{figure:12}
\end{center}
\end{figure}

\begin{table}
\caption{Variation in $\tau^{h}_{L}$, $\tau^{s}_{L}$, $\vartheta^{h}_{L}$, $\vartheta^{s}_{L}$ with chain 
length, in the Hubbard model with $\frac{U}{\mid t \mid}$=2.0,4.0,6.0, and the PPP model with $t$=-2.4 eV and 
$U$=11.26 eV.}
\begin{center}
\begin{tabular}{|c||c||c|c|c|}
\hline
Model Parameter & L               & 20 & 30 & 40  \\
\hline
& $\tau^{h}_{L}$ & 6.91679985 & 9.78119978 & 12.3353997 \\
& $\tau^{s}_{L}$ & 9.72179978 & 12.8567997 & 15.8069996 \\
$\frac{U}{\mid t \mid}$ = 2.0 & $\vartheta^{h}_{L}$ & 2.024057411 & 2.044738933 & 3.242699951 \\
& $\vartheta^{s}_{L}$ & 1.44006257 & 1.55559707 & 2.53052452 \\
& $\biggl(\vartheta^{h}_{L}/\vartheta^{s}_{L}\biggr)$ & 1.405534352 & 1.314439949 & 1.281433918 \\
\hline
& $\tau^{h}_{L}$ & 7.02239984 & 10.0517998 & 13.0811997 \\
& $\tau^{s}_{L}$ & 12.7445997 & 17.1599996 & 19.0409996 \\
$\frac{U}{\mid t \mid}$ = 4.0 & $\vartheta^{h}_{L}$ & 5.696058457 & 5.969080283 & 6.115647023 \\
& $\vartheta^{s}_{L}$ & 3.138584258 & 3.496503578 & 4.201460096 \\
& $\biggl(\vartheta^{h}_{L}/\vartheta^{s}_{L}\biggr)$ & 1.814849623 & 1.707156921 & 1.455600406 \\
\hline
& $\tau^{h}_{L}$ & 7.18739984 & 10.2827998 & 12.9755997 \\
& $\tau^{s}_{L}$ & 16.4669996 & 18.9815996 & 19.9319996 \\
$\frac{U}{\mid t \mid}$ = 6.0 & $\vartheta^{h}_{L}$ & 2.782647473 & 2.917493346 & 3.082709156 \\
& $\vartheta^{s}_{L}$ & 1.214550342 & 1.580477970 & 2.006823239 \\
& $\biggl(\vartheta^{h}_{L}/\vartheta^{s}_{L}\biggr)$ & 2.291092742 & 1.845956351 & 1.536113942 \\
\hline
& $\tau^{h}_{L}$ & 1.81499996 & 2.58059994 & 3.45179992 \\
& $\tau^{s}_{L}$ & 3.76859992 & 5.44499988 & 6.99599984 \\
$\text{PPP}$ & $\vartheta^{h}_{L}$ & 11.019283989 & 11.625203711 & 11.588157172 \\
& $\vartheta^{s}_{L}$ & 5.307010674 & 5.509641995 & 5.717553018 \\
& $\biggl(\vartheta^{h}_{L}/\vartheta^{s}_{L}\biggr)$ & 2.076363638 & 2.109974427 & 2.026768643 \\
\hline
\end{tabular}
\end{center}
\end{table}

\begin{figure}
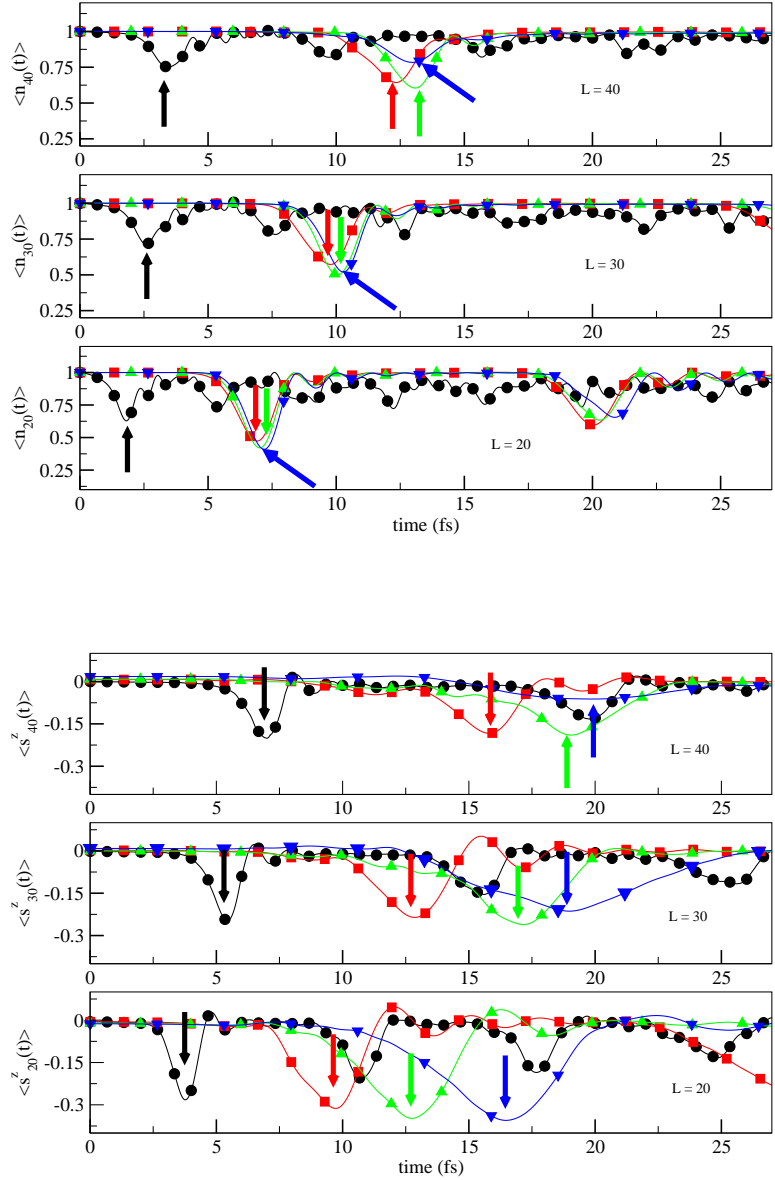

\begin{center}
\vspace*{-2.5cm}
\epsfig{file=chrgL_hubvsPPP.eps, width = 4.0 in} \\
\vspace*{1.5cm} \epsfig{file=spinL_hubvsPPP.eps, width = 4.0 in}
\caption{(Color online)
Comparison of the time evolution profiles of $\bra n_{L}(t) \ket$ (top curve) and $\bra s^{z}_{L}(t) \ket$ 
(bottom curve) for regular PPP chains with those of regular Hubbard chains, for chain lengths of 20, 30 and 
40 sites. The color coding with symbols, is as follows: black curve with circles: PPP model; red curve with 
squares=Hubbard model with $\frac{U}{\mid t \mid}$=2.0; green curve with up triangles=Hubbard model with 
$\frac{U}{\mid t \mid}$=4.0; blue curve with down triangles=Hubbard model with $\frac{U}{\mid t \mid}$=6.0. 
The position of $\tau^{h/s}_{L}$ is shown with the aid of arrows: PPP model (black arrow), Hubbard model with
$\frac{U}{\mid t \mid}$=2.0 (red arrow), 4.0 (green arrow), and 6.0 (blue arrow).}
\label{figure:13}
\end{center}
\end{figure}

\newpage
\section{SUMMARY AND OUTLOOK} 

We have developed a time-dependent DMRG scheme called DTWT technique by combining the salient features of 
the LXW and TST algorithms. This time-dependent DMRG technique is free from the drawbacks associated with 
both the parent methods while possessing their strengths. Our scheme is faster and more accurate than both 
the LXW and TST methods. The TST technique targets one time step of length $\Delta \tau$ and evolves the 
wave packet $\mid \psi(t) \ket$ over this time step, using four states constructed over a time interval 
$\Delta t$ which is larger than $\Delta \tau$. The time step $\Delta t$ is used for constructing the Hilbert 
space representing the time propagating wave packet. However, our studies have revealed that this usually 
short time-step is unsuitable for representing the Hilbert space of time-evolving system, the reason being, 
$\Delta t$ has poor information about the {\it future} states, along the trajectory of time propagation of 
the initial wave packet. Hence, the adaptively constructed Hilbert space of the time evolving wave function 
fails to follow the evolution successfully. The LXW scheme targets the whole time evolution interval at 
every system size of the infinite DMRG algorithm. Hence, even though the dimension of DMEV basis ({\it m}) is
fixed, the adaptively built Hilbert space at every system size successfully follows the time evolving 
wave function. However, the DMEV cut-off is significantly less than the number of retained target states. 
Hence the LXW procedure is capable of constructing the desired Hilbert space successfully. However, the time 
evolution of the final system is inaccurate due to the fact that number of target states is usually much
larger than the DMRG cut-off. Furthermore, the LXW scheme applied to a finite-system size, is not quasiexact.
DTWT circumvents these problems efficiently by considering a double time window (2$\Delta t$) for basis 
adaptation and single time window ($\Delta t$) which is embedded within the former time window, for 
evolution. The extra $p\Delta t$ steps within a double time window which are used for targeting, ensures that 
for every single time window propagation, the basis gains the information about the next single time window. 
This ensures higher accuracy of DTWT technique compared to both the LXW and the TST schemes.

We have found that this method is applicable not only to chains but also to lattices with other topologies 
like rings and ring-chain systems. Using this technique we have performed non-equilibrium dynamics of 
spin-charge transport in the PPP model which harbors long-range Coulomb repulsion. Real-time dynamics of 
spin and charge transport in these systems is still in its infancy in the literature. In future we intend to 
address the effect of dimerization on the spin-charge separation phenomena as well as, study spin and charge 
transport in Y-junctions. Using our DTWT procedure, we intend to study non-equilibrium transport in a 
molecule between two leads (having lead-molecule-lead geometry). 

Further improvements of the time evolution in the DTWT scheme can be carried out. We recognize that  the
time-dependent DMRG schemes have two sources of error: (i) the truncation error due to Hilbert-space 
truncation and (ii) time propagation error due to finite time step employed in the 
numerical solution of the TDSE. Besides, there is also the problem of stability of the numerical method 
for time steps larger than a critical value. We intend to overcome the latter two problems by using the 
Chebyshev polynomial-based decomposition of the time evolution operator.
Exact propagation of the wave packet $\mid \psi(t) \ket$ using
\beq
\mid \psi(t+\Delta t) \ket ~=~ \sum_{n=1}^{M} c_{n} e^{-i\Delta t(E_{n}-E_{1})} \mid \phi_{n}(t) \ket,
\eeq
where $\mid \psi(t) \ket$ = $\sum_{n=1}^{M} c_{n} \mid \phi_{n}(t) \ket$, $\mid \phi_{n}(t) \ket$ is the
{\it n}th eigenstate of $H_{\text{DMRG}}$, $E_{1}$ is the ground state energy, $E_{n}$ is the {\it n}th 
eigen-energy, $M$ is the order of the matrix $H_{\text{DMRG}}$, and $c_{n}$ = $\bra \phi_{n} \mid \psi(t) 
\ket$, has no error associated with solving the TDSE. Chebyshev expansion of the time evolution operator 
\cite{kosloff1,lubich} has the same advantages as exact expansion of the time evolution operator. The 
Chebyshev expansion of the state $\mid \psi(t+\Delta t) \ket$ is given by
\beq
\begin{split}
\mid \psi(t+\Delta t) \ket &=~ e^{-iH\Delta t} \mid \psi(t) \ket \\
                           &\approx~ \sum_{n=0}^{M} a_{n} T_{n}(\mathbbm{H}) \mid \psi(t) \ket, 
\end{split}
\eeq
where $T_{n}(\mathbbm{H})$ is the {\it n}th Chebyshev polynomial of the first kind, $\mathbbm{H}$ is the 
scaled Hamiltonian with eigenvalues ranging from $[-1.0, 1.0]$, $a_{n}$ is coefficient of 
$T_{n}(\mathbbm{H})$ given by
\beq
a_{n} = (2-\delta_{n0})e^{-i\Delta t\frac{(E_{max}+E_{min})}{2}} (-i)^{n} J_{n}(\Delta t\frac{E_{max}-E_{min}}{2}).
\eeq   
$E_{min}$ and $E_{max}$ are the minimum and maximum energies of $H$; $J_{n}$ is the {\it n}th
order Bessel function of the first kind. The Chebyshev expansion of the evolution operator can be 
evaluated up to machine accuracy and there will be no error associated with this time propagation 
scheme for any arbitrary step-size. The use of Chebyshev polynomial expansion scheme is expected to 
render the DTWT technique free from any time step error besides decreasing the computational time.

\begin{center}
{\bf ACKNOWLEDGEMENTS}  
\end{center}
T.D. acknowledges S. R. White for helpful suggestions during the implementation of the TST 
algorithm, Diptiman Sen for his comments and suggestions with the results and discussion part.
This work was supported by DST India and the Swedish Research Link Program under the Swedish Research
Council.

\end{document}